%% file: arxiv.tex
\documentclass[a4paper,UKenglish,cleveref, autoref,numberwithinsect,thm-restate]{lipics-v2021}

\pdfoutput=1 
\hideLIPIcs

\graphicspath{{./graphics/}}

\title{Feedback Vertex Set, Max Cut, and Edge Coloring on the Strong Product of a Graph and a Path}

\titlerunning{FVS, MC, and EC on the Strong Product of a Graph and a Path}

\author{Jan Diester}{Karlsruhe Institute of Technology, Germany}{}{}{}

\author{Kolja Kühn}{Karlsruhe Institute of Technology, Germany}{kolja.kühn@kit.edu}{https://orcid.org/0009-0006-0746-6876}{Supported by KiKIT — the Helmholtz pilot program for core informatics research at the KIT.}

\author{Manuel Rapp}{Karlsruhe Institute of Technology, Germany}{}{}{}

\author{Samuel Schneider}{Karlsruhe Institute of Technology, Germany}{samuel.schneider@kit.edu}{https://orcid.org/0009-0002-9680-4048}{}

\author{Torsten Ueckerdt}{Karlsruhe Institute of Technology, Germany}{torsten.ueckerdt@kit.edu}{https://orcid.org/0000-0002-0645-9715}{Funded by the Deutsche Forschungsgemeinschaft (DFG, German Research Foundation) – 532213793}

\authorrunning{J. Diester, K. Kühn, M. Rapp, S. Schneider, T. Ueckerdt}

\Copyright{Jan Diester, Kolja Kühn, Manuel Rapp, Samuel Schneider, Torsten Ueckerdt}



\category{}

\relatedversion{}

\nolinenumbers

\usepackage{xspace}
\usepackage{xcolor}
\usepackage{complexity}

\newcommand{\ceil}[1]{\left\lceil #1 \right\rceil}
\newcommand{\floor}[1]{\left\lfloor #1 \right\rfloor}
\newcommand{\alternating}{alternating}
\newcommand{\project}[2]{\ensuremath{#1.#2}}
\newcommand{\fvs}{\textsc{Feedback Vertex Set}\xspace}
\newcommand{\wfvs}{\textsc{Weighted Feedback Vertex Set}\xspace}

\newcommand{\vc}{\textsc{Vertex Cover}\xspace}
\newcommand{\mc}{\textsc{Max Cut}\xspace}
\newcommand{\ec}{\textsc{3-Edge Coloring}\xspace}
\newcommand{\minUncut}{\textsc{Min Uncut}\xspace}

\newtheorem{question}[theorem]{Question}
\crefname{observation}{Observation}{Observations}
\crefname{question}{Question}{Questions}

\definecolor{signalblue1}{RGB}{58,118,240}
\DeclareTextFontCommand{\emph}{\color{signalblue1}\em}

\let\oldP\P
\renewcommand{\P}{\oldP\xspace}
\let\oldFPT\FPT
\renewcommand{\FPT}{\oldFPT\xspace}
\renewcommand{\Box}{\mathbin{\square}}

\DeclareCaptionLabelFormat{bold}{\textbf{\sffamily(#2)}}
\input{table}

\begin{document}

\maketitle

\begin{abstract}
    We examine the computational complexity of well-known problems on subgraphs of the strong product of two paths.
    Namely, for \textsc{Max Cut} we show \NP-hardness on induced subgraphs of $P \boxtimes P$ answering a question by Dvo{\v{r}}{\'a}k, Huynh, Joret, Liu, and Wood (2021).
    For \textsc{3-Edge Coloring} we show \NP-hardness on subgraphs of $P \boxtimes P$ and that it is equivalent to \textsc{3-Edge Coloring} on planar graphs for induced subgraphs of $P \boxtimes P$.
    It is a long standing open question whether \textsc{3-Edge Coloring} on planar graphs is \NP-hard.
    
    As a second contribution, we consider graphs of the form $G \boxtimes P$ and show that \textsc{Feedback Vertex Set} on these graphs is \FPT in the clique-width of $G$.
    Furthermore, we show that it can be solved in polynomial time if $G$ is chordal or co-comparable and that the computational complexity depends on the parity of $P$ if $G$ is bipartite.
    Namely, it is solvable in polynomial time if $P$ has an even number of vertices but is \NP-complete otherwise.
    For \textsc{Max Cut}, we show that it can be solved in polynomial time if $G$ is bipartite.
\end{abstract}

\section{Introduction}
\label{sec:introduction}

Graph products have been a popular subject to study for many decades, e.g., spanning large parts in several textbooks~\cite{graph_products_handbook,graph_coloring_book}.
A graph product is a binary (not necessarily symmetric) operator $\circ$ that takes two graphs $G,H$ (the factors) and gives a resulting graph $G \circ H$.
There are several variants, including the Cartesian product, tensor product, lexicographic product, strong product, modular product, rooted product, and zig-zag product, just to name a few.

In the last few years, the strong product has received a lot of attention due to its central role in the \emph{Graph Product Structure Theory}.
The starting point was a result~\cite{planar_bounded_queuenumber} stating that for every planar graph $G$ there exists a graph $H$ of treewidth at most~$8$\footnote{The bound on the treewidth was later improved to $6$~\cite{improved_planar_product_structure}.} such that $G \subseteq H \boxtimes P$, i.e., $G$ is a subgraph of the strong product of $H$ and a path $P$.
Crucially, the planar graph $G$ itself can have arbitrarily large treewidth.
So, in a way, the planar product structure theorem asserts that one can reduce the complexity of any planar graph $G$ by ``factoring out'' a single path.
Here, the complexity is given by its treewidth, which is a graph parameter that is long known to be particularly useful for designing efficient algorithms~\cite{courcelles_theorem}.

Consequently, researchers wondered whether product structure can be exploited algorithmically.
In particular, for any \NP-complete problem and any concrete concept of complexity, one can ask the following.

\begin{question}
    \label{que:general}
    Assume one is given a factorization of a graph $G$ into a graph $H$ (of ``low complexity'') and a path $P$, i.e., $G \subseteq H \boxtimes P$.
    Can one solve a problem for $G$ efficiently, provided that the problem is easy to solve on $H$?
\end{question}

For example, Dvo{\v{r}}{\'a}k, Huynh, Joret, Liu, and Wood~\cite{question_MaxCut} explicitly ask whether \mc can be solved efficiently on subgraphs of $H \boxtimes P$ when $H$ has bounded treewidth.
We answer this in the negative by showing that \mc is \NP-complete already for subgraphs of $P \boxtimes P$, i.e., subgraphs of the strong grid.
(We discuss our results in detail below.)

Of course, \cref{que:general} is too generic to have a definite answer.
It crucially depends on the considered decision problem (and the notion of complexity).
In this paper we focus on the following three fundamental \NP-hard problems.

\begin{itemize}
    \item The \fvs problem asks for a given graph $G = (V,E)$ and integer $k$, whether there exists a set of vertices $U \subseteq V$ with $|U| \leq k$ such that $G - U$ is a forest.
    \item The \mc problem asks for a given graph $G = (V,E)$ and integer $k$, whether there exists a set of vertices $U \subseteq V$ such that $|\{ uv \in E \colon u \in U, v\notin U\}| \geq k$.
    \item The \ec problem asks for a given graph $G = (V,E)$, whether there exists a coloring $c \colon E \to \{1,2,3\}$ such that for any two distinct edges $e_1,e_2 \in E$ with a common endpoint it holds $c(e_1) \neq c(e_2)$.
\end{itemize}

Each of the above problems is known to be efficiently solvable on forests and, more generally, on graphs of bounded treewidth.
On planar graphs, \fvs is known to be \NP-complete~\cite{garey_johnson}, while \mc is known to be polynomial-time solvable~\cite{mc_planar_graphs}.
The computational complexity of the \ec problem on planar graphs is a prominent open problem in the field.
In combination with the planar product structure theorem, this makes these three problems particularly interesting to us.
But we also discuss some further decision problems in our concluding remarks in \cref{sec:conclusion}.

\subparagraph*{Our Results.}

We consider each of \fvs, \mc, and \ec in three scenarios: for subgraphs of $G \boxtimes P$, for induced subgraphs of $G \boxtimes P$, and for the full graph $G \boxtimes P$.
In the most general setting of arbitrary subgraphs of $G \boxtimes P$, all three problems turn out to be \NP-complete, already when $G$ is a path.

\begin{theorem}
    \label{thm:main-thm-subgraphs}
    Each of the following holds.
    \begin{enumerate}
        \item \fvs is \NP-complete on subgraphs of $P \boxtimes P$~\cite{fvs_induced_grid}.
        \item \mc is \NP-complete on subgraphs of $P \boxtimes P$.
        \item \ec is \NP-complete on subgraphs of $P \boxtimes P$.
    \end{enumerate}
\end{theorem}

For the more restricted setting of induced subgraphs of $P \boxtimes P$, \fvs and \mc remain \NP-complete, while \ec turns out to be polynomially equivalent to \ec on planar graphs.

\begin{theorem}
    \label{thm:main-thm-induced-subgraphs}
    Each of the following holds.
    \begin{enumerate}
        \item \fvs is \NP-complete on induced subgraphs of $P \boxtimes P$~\cite{fvs_induced_grid}.
        \item \mc is \NP-complete on induced subgraphs of $P \boxtimes P$.
        \item \ec on induced subgraphs of $P \boxtimes P$ is polynomially equivalent to \ec on planar graphs.
    \end{enumerate}
\end{theorem}

In our most restricted setting, the input is a graph $G$ and an integer $\ell$ and the task is to solve the problem on $G \boxtimes P_\ell$, i.e., the strong product of $G$ with a path on $\ell$ vertices.
It is not hard to see that for every $\ell \geq 2$ the full product $G \boxtimes P_\ell$ admits a proper $3$-edge-coloring if and only if $G \boxtimes P_\ell$ has maximum degree at most~$3$.
Thus, the \ec problem on $G \boxtimes P_\ell$ becomes trivial for every $\ell \geq 2$.

The \mc problem on the full product $G \boxtimes P_\ell$ seems to be a non-trivial problem, even for graphs $G$ from very restricted graph classes.
We show that for every $\ell \in \mathbb{N}$ \mc on $G \boxtimes P_\ell$ is solvable in linear time on bipartite graphs $G$.
However, the computational complexity of \mc on $G \boxtimes P_\ell$ remains open even for cactus graphs $G$.

The \fvs problem on $G \boxtimes P_\ell$ shows an interesting behavior depending on the parity of $\ell$.
If $\ell$ is even, then \fvs on $G \boxtimes P_\ell$ is equivalent to the \vc problem on $G$.
So in particular, if $G$ is a subcubic graph, then \fvs on $G$ is polynomial-time solvable~\cite{fvs_max_degree_3}, while \fvs on $G \boxtimes P_\ell$ is \NP-complete, since \vc on subcubic graphs is known to be \NP-complete~\cite{vc_cubic_graphs}.

If $\ell$ is odd, then \fvs on $G \boxtimes P_\ell$ is equivalent to a problem on $G$ that we call $\ell$-weighted \fvs.
We show
that if \fvs is \NP-hard on a graph class $\mathcal{G}$ that is closed under so-called $k$-hair extensions, then \fvs is also \NP-hard on $G \boxtimes P_\ell$ with $G \in \mathcal{G}$ and odd $\ell$.
As a consequence, for the class $\mathcal{G}$ of all planar bipartite graphs, \fvs on $G \boxtimes P_\ell$ with $G \in \mathcal{G}$ is \NP-hard for every odd $\ell \in \mathbb{N}$, but efficiently solvable for every even~$\ell \in \mathbb{N}$.

\begin{theorem}
    \label{thm:main-thm-full-product}
    Each of the following holds.
    \begin{enumerate}
        \item For every $\ell \in \mathbb{N}$ \fvs on $G \boxtimes P_\ell$ is polynomial-time solvable for chordal graphs $G$, for co-comparability graphs $G$, and for graphs $G$ of bounded clique-width.
        \item For every even $\ell \in \mathbb{N}$ \fvs on $G \boxtimes P_\ell$ is polynomially equivalent to \vc on $G$.
        \item For every odd $\ell \in \mathbb{N}$ \fvs on $G \boxtimes P_\ell$ is polynomially equivalent to $\ell$-weighted \fvs on $G$.
        \item For every $\ell \in \mathbb{N}$ \mc on $G \boxtimes P_\ell$ is linear-time solvable for bipartite graphs $G$.
        \item For every $\ell \in \mathbb{N}$, $\ell \geq 2$, \ec on $G \boxtimes P_\ell$ is linear-time solvable.
    \end{enumerate}
\end{theorem}

See \cref{fig:overview} for an overview of our results.

\begin{table}
    \centering
    \resultstable
    \caption{
        The computational complexity of \fvs, \mc, and \ec on subgraphs of $P \boxtimes P$, induced subgraphs of $P \boxtimes P$, and the full product $G \boxtimes P$.
        \textsc{VC} refers to \vc, \textsc{$\ell$-FVS} refers to $\ell$-weighted \fvs, and \textsc{$3$-EC} refers to \textsc{$3$-Edge Coloring}.
        Cells listing a problem $Q$ mean that in this case the computational complexity is equivalent to that of $Q$.
    }
    \label{fig:overview}
\end{table}

\subparagraph*{Organization of the paper.}
In \cref{sec:preliminaries} we give the formal definition of the relevant graph products and central notions used throughout this paper.
We prove our results on \fvs in \cref{sec:feedback-vertex-set}.
In \cref{sec:grid-embeddings} we show that every subcubic graph $G$ admits a subdivision $G'$ (of polynomial size) that is a subgraph of the strong grid $P \boxtimes P$.
If $G$ is additionally planar, then $G'$ is even an induced subgraph of $P \boxtimes P$.
We use these subdivisions to show our hardness for \mc in \cref{sec:max-cut} and \ec in \cref{sec:edge-coloring}.
We conclude with a discussion and some open problems in \cref{sec:conclusion}.
Omitted material and proofs are provided in the Appendix.
This includes the polynomial-time algorithm for \mc on the full product $G \boxtimes P_\ell$ when $G$ is bipartite.

\subparagraph*{Use of AI.}
None of our main results were obtained using AI.
We solely used AI to find the example in \cref{fig:counterexample_alternating_maxcut} that certifies that our approach for \mc on $G \boxtimes P_\ell$ does not extend to all graphs $G$.
We did not use AI in any form in the write-up of this paper.


\section{Preliminaries}
\label{sec:preliminaries}

\begin{figure}
    \centering
    \includegraphics{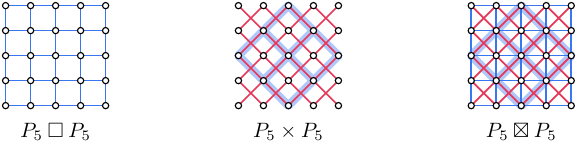}
    \caption{The Cartesian, tensor, and strong product of two paths on five vertices.
    Both the tensor and the strong product contain an induced $P_3 \Box P_3$ subgraph highlighted in blue.}
    \label{fig:prel:products}
\end{figure}

All graphs considered in this paper are simple and undirected.
A graph $G$ is called \emph{planar} if it can be drawn in the plane without any edge crossings.
For two graphs $G_1 = (V_1, E_1), G_2 = (V_2, E_2)$ we define the \emph{Cartesian product} $G_1 \Box G_2$, the \emph{tensor product} $G_1 \times G_2$ and the \emph{strong product} $G_1 \boxtimes G_2$ as graphs on vertex set $V_1 \times V_2$ and the following edge sets:

Two vertices $(u_1, u_2), (v_1, v_2)$ are adjacent in $G_1 \Box G_2$ if $u_1 = v_1 \text{ and } u_2v_2 \in E_2$, or $u_1v_1 \in E_1 \text{ and } u_2 = v_2$.
They are adjacent in $G_1 \times G_2$ if $u_1v_1 \in E_1 \text{ and } u_2v_2 \in E_2$.
Finally, the edge set of $G_1 \boxtimes G_2$ is the union of the edge sets of $G_1 \Box G_2$ and $G_1 \times G_2$.
\cref{fig:prel:products} shows the Cartesian, tensor, and strong product of two path graphs.

For an integer $n \geq 1$, we denote by $P_n$ the path on $n$ vertices.
We refer to $P_n \Box P_n$ as the \emph{$n \times n$ square grid} and to $P_n \boxtimes P_n$ as the \emph{$n \times n$ strong grid}.
Further, we simply use $P \Box P$ and $P \boxtimes P$ whenever the concrete size of the grid is not relevant, i.e., we just need the grid to be large enough.
For example, we may say that a graph $G$ is a subgraph or induced subgraph of $P \Box P$ or $P \boxtimes P$ without specifying the length of the two paths.
(If $n = |V(G)|$, then $P_n \Box P_n$ and $P_n \boxtimes P_n$ are already large enough.)

By definition, $P \boxtimes P = (P \Box P) \cup (P \times P)$, i.e., the strong grid is the edge-disjoint union of $P \times P$ and the square grid $P \Box P$.
We call the edges of $P \boxtimes P$ in $P \times P$ \emph{diagonal} and the edges of $P \boxtimes P$ in $P \Box P$ \emph{orthogonal}.
Clearly, $P \Box P$ is bipartite with a unique bipartition $(W,B)$, which we can imagine as the white and black vertices, respectively.
Every diagonal edge in $P \boxtimes P$ has both endpoints in $W$ or both endpoints in $B$.
In fact, the induced subgraph of $P \times P$ on vertex-set $W$ (and symmetrically $B$) is isomorphic to an induced subgraph of the square grid $P \Box P$ sitting rotated by 45 degrees (and stretched by a factor of $\sqrt{2}$) in $P \boxtimes P$.
See \cref{fig:prel:products} for an illustration.
Crucially, $P_n \Box P_n$ is an induced subgraph of $P_{2n} \boxtimes P_{2n}$.

\begin{observation}
    \label{obs:grid-to-induced-strong-grid}
    Every induced subgraph of $P_n \Box P_n$ is an induced subgraph of $P_{2n} \boxtimes P_{2n}$ consisting only of diagonal edges.
\end{observation}

\section{Feedback Vertex Set}
\label{sec:feedback-vertex-set}

In this section, we examine the \emph{\fvs problem} that asks for a given graph $G$ and an integer~$k$ whether there is a set $F\subseteq V(G)$ with $|F|\leq k$ such that $G - F$ is a forest.
\fvs is known to be \NP-hard
on induced subgraphs of $P \Box P$~\cite{fvs_induced_grid} and, thus, also on induced subgraphs of $P \boxtimes P$ by \cref{obs:grid-to-induced-strong-grid}.
In contrast, we show for the full product $G \boxtimes P$ that \fvs is in \P for graphs $G$ with bounded clique-width.
We do so by showing that for every $\ell \in \mathbb{N}$ finding a feedback vertex set on $G \boxtimes P_{\ell}$ is equivalent to finding a weighted feedback vertex set on $G\boxtimes P_2$.

Additionally, we show that the computational complexity of \fvs on $G \boxtimes P_\ell$ may differ depending on the parity of $\ell$.
Namely, for even $\ell$, we show in \cref{se:fvs_even_length} that \fvs on $G\boxtimes P_\ell$ is equivalent to \vc on $G$ (cf.\ \cref{cor:vertex_cover}).
For odd $\ell$, we show in \cref{se:fvs_odd_length} that finding a feedback vertex set on $G\boxtimes P_\ell$ is equivalent to finding a maximum induced forest in $G$, where isolated vertices contribute more weight than non-isolated vertices (cf.\ \cref{obs:ell_weighted_forest}).
As a consequence, we get for the class $\mathcal{G}$ of all bipartite graphs and all $\ell \in \mathbb{N}$ that \fvs on $G \boxtimes P_\ell$ for graphs $G \in \mathcal{G}$ is in \P for even $\ell$ and \NP-hard for odd $\ell$ (cf.\ \cref{thm:fvs_hardness_odd}).

\smallskip

Let $G$ be a graph and $P_\ell = (p_1,\dots, p_\ell)$ be a path.
For $i \in [\ell]$ we set $V_i = V(G) \times \{p_i\}$ and $G_i = (G \boxtimes P_\ell)[V_i]$, and call $G_i$ the \emph{$i$-th copy of~$G$}.
I.e., $G_i$ is the subgraph of $G \boxtimes P_\ell$ induced by the vertices in $V_i$.
For a set of vertices $A \subseteq V(G \boxtimes P_\ell)$ we define the set of vertices in $A$ projected onto~$G$ as $\project{A}{G} = \{ v \in V(G) \mid \exists i \in [\ell] \colon (v,p_i) \in A\}$.
The set $A$ is \emph{\alternating} if for each $i \in [\ell -2]$ we have $\project{(A \cap V_i)}{G}=\project{(A \cap V_{i+2})}{G}$.
That is, $A$ is alternating if $A$ contains the same vertices in every even copy of~$G$, and the same vertices in every odd copy of~$G$.

We start by proving a technical lemma that asserts properties of a (minimum) feedback vertex set of $G \boxtimes P_2$ that we may wish for in later proofs.
Roughly speaking, given any feedback vertex set $F$ of $G \boxtimes P_2$, ``pushing'' every vertex in $F$ that is in $G_1$ to $G_2$ results again in a feedback vertex set.
See \cref{fig:fvs_p2} for an example.

\begin{lemma}\label{le:fvs_p_2}
    For every graph $G$ and every feedback vertex set $F$ of $G \boxtimes P_2$, there is a feedback vertex set $F'$ of $G \boxtimes P_2$ such that each of the following holds.
    \begin{enumerate}[(a)]
        \item $|F'| = |F|$\label{le:fvs_p_2_cardinality}
        \item $|F' \cap V_1| \leq |F \cap V_1|$\label{le:fvs_p_2_dominates}
        \item $\project{(F' \cap V_1)}{G} \subseteq \project{(F' \cap V_2)}{G}$\label{le:fvs_p_2_subset}
        \item For every $\ell \geq 2$, the unique alternating set $F''$ on $G \boxtimes P_\ell$ with $F'' \cap (V_1 \cup V_2) = F'$ is a feedback vertex set of $G \boxtimes P_\ell$.\label{le:fvs_p_2_f''}
    \end{enumerate}
\end{lemma}
\begin{proof}
    Given $F$, we start by constructing the feedback vertex set $F'$ and then show that it has the desired properties.
    To this end, consider the sets $F_1 = \project{(F \cap V_1)}{G}$ and $F_2 = \project{(F \cap V_2)}{G}$, and define $F' = ((F_1 \cup F_2) \times \{p_2\}) \cup ((F_1 \cap F_2) \times \{p_1\})$.

    \begin{figure}
        \centering
        \includegraphics{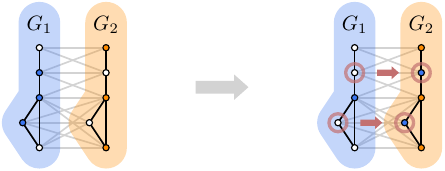}
        \caption{On the left is a graph $G \boxtimes P_2$ with a feedback vertex set $F$ consisting of the colored vertices.
        The sets $F_1$ and $F_2$ correspond to the colored vertices in $G_1$ and $G_2$, respectively.
        On the right is the same graph with the corresponding feedback vertex set $F'$.
        The colored vertices in $G_1$ form the set $(F_1 \cap F_2) \times \{p_1\}$.
        The colored vertices in $G_2$ form the set $(F_1 \cup F_2) \times \{p_1\}$.}
        \label{fig:fvs_p2}
    \end{figure}

    First, note that indeed $|F'| = |F|$ as $|F_1 \cup F_2| + |F_1 \cap F_2| = |F|$.
    Furthermore, by the definition of $F'$ we have $|F' \cap V_1| \leq |F \cap V_1|$ and $\project{(F' \cap V_1)}{G} \subseteq \project{(F' \cap V_2)}{G}$.
    To see that $F'$ is a feedback vertex set of $G \boxtimes P_2$, note that for each vertex $u \in V(G)$ there are the same number of copies of it in $F$ as in $F'$.
    Thus, since for each vertex $u \in V(G)$, the vertices $(u,p_1)$ and $(u,p_2)$ are twins in $G \boxtimes P_2$, i.e., $N[(u,p_1)] = N[(u,p_2)]$, it holds that $G \boxtimes P_2 - F$ is isomorphic to $G \boxtimes P_2 - F'$.

    Finally, let $\ell \geq 2$ and $F''$ be the unique alternating set on $G \boxtimes P_\ell$ with $F'' \cap (V_1 \cup V_2) = F'$.
    To show that $F''$ is a feedback vertex set of $G \boxtimes P_\ell$, we consider each component $K$ of $(G \boxtimes P_\ell) - F''$, distinguishing whether $K$ contains some vertex in an even copy of $G$ or not.
    
    The definition of $F'$ and $F''$ imply for every $u \in V(G)$ that if $(u,p_i) \in F''$ for some odd $i \in [\ell]$, then $(u,p_i) \in F''$ for all $i \in [\ell]$.
    In other words, if $(u,p_i) \notin F''$ for some even $i \in [\ell]$, then $(u,p_i) \notin F''$ for all $i \in [\ell]$.
    Hence, if component $K$ of $(G \boxtimes P_\ell) - F''$ contains some vertex $(u,p_i)$ in an even copy of $G$, then $K$ contains the entire path $\{u\} \boxtimes P_\ell$.
    As $(G \boxtimes P_2) - F'$ contains no triangles, this implies that for every edges $uw$ of $G$ we have $(\{w\} \boxtimes P_\ell) \cap K = \emptyset$.
    Thus $K = \{u\} \boxtimes P_\ell$ is a path.

    On the other hand, if component $K$ of $(G \boxtimes P_\ell) - F''$ contains no vertex of any even copy of $G$, then $K$ is completely contained in some odd copy of $G$.
    But then $K$ is isomorphic to a component of $(G \boxtimes P_2) - F'$ and hence a tree, since $F'$ is a feedback vertex set of $G \boxtimes P_2$.
\end{proof}

From \cref{le:fvs_p_2} we deduce that $G\boxtimes P_\ell$ has a minimum feedback vertex set that is alternating.
The second part of the statement is needed later for the case that $\ell$ is odd in \cref{se:fvs_odd_length}.

\begin{lemma}\label{le:fvs_alternating}
    For every graph $G$ and $\ell \geq 2$, there is a minimum feedback vertex set $F$ of $G \boxtimes P_\ell$ that is alternating.
    Furthermore, there is such a set $F$ with $\project{(F \cap V_1)}{G} \subseteq \project{(F \cap V_2)}{G}$.
\end{lemma}
\begin{proof}
    Let $F$ be a minimum feedback vertex set of $G \boxtimes P_\ell$.
    Our goal is to construct an alternating feedback vertex set $F'$ with $|F'| \leq |F|$.
    To do this, let $G_i$ and $G_{i+1}$ be the two consecutive copies of $G$ that minimize the size of $V(G_i \cup G_{i+1}) \cap F$.
    We assume without loss of generality that $i$ is odd.
    If $\ell$ is even there is an alternating feedback vertex set of the same size as $F$ by \cref{le:fvs_p_2} (choose $V(G_i \cup G_{i+1}) \cap F$ as the feedback vertex set of $G \boxtimes P_2$).
    The second part of the statement then follows immediately from \cref{le:fvs_p_2}\labelcref{le:fvs_p_2_subset}.

    Consider now that $\ell$ is odd.
    Let $F''$ be the alternating feedback vertex set of $G \boxtimes P_{\ell}$ given by \cref{le:fvs_p_2} by choosing $V(G_i \cup G_{i+1}) \cap F$ as the feedback vertex set of $G \boxtimes P_2$.
    We claim that $|F''| \leq |F|$, i.e., $F''$ is a minimum feedback vertex set.
    Note that once again, the second part of the statement would then follow immediately from \cref{le:fvs_p_2}\labelcref{le:fvs_p_2_subset}.

    As $\ell \geq 3$, we may assume that $i < \ell$.
    Then $i+2 \leq \ell$ and $P_\ell - p_{i+2}$ is the union of two (possibly empty) paths, each with an even number of vertices.
    Thus we can partition (the vertices of) $G \boxtimes P_\ell$ into $G_{i+2}$ and $\lfloor \frac{\ell}2 \rfloor$ copies of $G \boxtimes P_2$.
    We count the total number of vertices in $F''$ by counting in each part.
    
    First, note that $|V_{i+2} \cap F| \geq |V_i \cap F|$ by the choice of $i$, and $|V_{i+2} \cap F''| \leq |V_i \cap F|$ by \cref{le:fvs_p_2}\labelcref{le:fvs_p_2_dominates}.
    Secondly, for each of the $\lfloor \frac{\ell}2 \rfloor$ copies $H$ of $G \boxtimes P_2$ we have that $|V(H) \cap F| \leq |V(G_i \cup G_{i+1}) \cap F|$ by the choice of $i$, and $|F'' \cap V(H)| \leq |F \cap V(H)|$ by \cref{le:fvs_p_2}\labelcref{le:fvs_p_2_cardinality}.
    Thus, 
    \[
        |F''| = |V_{i+2} \cap F''| + \sum_H |V(H) \cap F''| \leq |V_{i+2} \cap F| + \sum_H |V(H) \cap F| = |F|.\qedhere
    \]
\end{proof}

Using \cref{le:fvs_alternating}, we can interpret the \fvs problem of $G \boxtimes P_\ell$ as a weighted version of \fvs on $G \boxtimes P_2$ by assigning each vertex in $G_1 \subset G \boxtimes P_2$ weight $\ceil{\frac{\ell}{2}}$, i.e., the number of odd copies of $G$ in $G \boxtimes P_\ell$, and each vertex in $G_2 \subset G \boxtimes P_2$ weight $\floor{\frac{\ell}{2}}$, i.e., the number of even copies of $G$ in $G \boxtimes P_\ell$.

\begin{observation}\label{obs:weighted_fvs}
    Let $G$ be a graph, $P=(p_1,p_2)$ be a path on two vertices, $\ell \in \mathbb{N}$, and $w \colon V(G \boxtimes P_2) \rightarrow \mathbb{N}$ be a weight function such that $w((v,p_1)) = \floor{\frac{\ell}{2}}$ and $w((v,p_2))=\ceil{\frac{\ell}{2}}$ for each $v \in V(G)$.
    Then, the minimum weight of a feedback vertex set of $G \boxtimes P_2$ with regard to $w$ is the same as the size of a minimum feedback vertex set of $G \boxtimes P_\ell$.
\end{observation}

By \cref{obs:weighted_fvs}, for any $\ell \in \mathbb{N}$ we can solve \fvs on $G \boxtimes P_\ell$ by solving \wfvs on $G \boxtimes P_2$.
In particular, we get the following.

\begin{lemma}
    \label{lem:fvs_same_time_as_wfvs}
    Let $\mathcal{G}$ be a graph class such that for each graph $G$, the graph $G \boxtimes P_2$ is in $\mathcal{G}$ as well.
    Then, for graphs $G\in \mathcal{G}$ and every $\ell\in \mathbb{N}$, \fvs on $G \boxtimes P_\ell$ can be solved in the same time as \wfvs on $G$.
\end{lemma}

\cref{lem:fvs_same_time_as_wfvs} gives algorithms for \fvs on $G \boxtimes P_\ell$ for graph $G$ from several graph classes.
For example, the clique-width of $G \boxtimes P_2$ is the same as the clique-width of $G$ (if $G$ has at least one edge)~\cite[Corollary~1]{clique_width_strong_product}.%
\footnote{Gurski shows that the clique-width of the \emph{lexicographic product} $G \cdot H$ is the maximum of the clique-width of $G$ and $H$. It is easy to see that $G \boxtimes P_2 = G \cdot P_2$. 
Note that this does not hold for paths $P_\ell$ with $\ell > 2$.}
Furthermore, \wfvs can be solved in $2^{O(k)}\cdot n$ on graphs of clique-width~$k$~\cite[Theorem~4.6]{fvs_bounded_clique_width}.

\begin{theorem}\label{thm:fvs_cliquewidth}
    For an $n$-vertex graph $G$ of clique-width~$k$ there is an algorithm that computes a minimum feedback vertex set of $G \boxtimes P_\ell$ for any $\ell \in \mathbb{N}$ in time $2^{O(k)}\cdot n$.
\end{theorem}

Similarly, we can observe that for a chordal graph $G$, the graph $G \boxtimes P_2$ is chordal as well.%
For the sake of completeness we include a short proof of this well-known statement.

\begin{lemma}\label{le:chordal_p2}
    Let $G$ be a chordal graph. 
    Then, the graph $G \boxtimes P_2$ is a chordal graph as well.
\end{lemma}
\begin{proof}
    Gavril~\cite[Theorem 3]{chordal_graphs_intersecion} shows that a graph is chordal if and only if it is the intersection graph of subtrees of a tree.
    That is, for each chordal graph $G$ there is a tree $T$ such that each vertex of $G$ corresponds to a subtree of $T$ and two vertices in $G$ are adjacent if and only if the two subtrees share at least one vertex.
    Clearly, adding an additional copy of each subtree results in such an intersection representation of $G \boxtimes P_2$.
\end{proof}

Since \wfvs can be solved in polynomial time on chordal graphs~\cite[Corollary 2.3\footnote{%
The corollary only states the result for unweighted feedback vertex set. The paragraph before the corollary explains the generalization to the weighted case.%
}]{fvs_chordal}, we get the following from \cref{lem:fvs_same_time_as_wfvs,le:chordal_p2}.

\begin{theorem}\label{thm:fvs_chordal}
    For a chordal graph $G$ there is an algorithm that computes a minimum feedback vertex set of $G \boxtimes P_\ell$ for any $\ell \in \mathbb{N}$ in polynomial time.
\end{theorem}

Finally, if $G$ is a co-comparability graph, then so is $G \boxtimes P_2$.
For the sake of completeness we include a short proof of this well-known statement.

\begin{lemma}\label{le:cocomp_p2}
    If $G$ is a co-comparability graph, then $G \boxtimes P_2$ is a co-comparability graph.
\end{lemma}
\begin{proof}
    A graph $G$ is a co-comparability graph if and only if $G$ is a function graph~\cite[Theorem 1]{cocomparability_graphs_intersection}, i.e., an intersection graph of graphs of continuous functions $f_i \colon [0,1] \rightarrow \mathbb{R}$.
     Adding another copy of each function results in such an intersection representation of~$G \boxtimes P_2$.
\end{proof}

Since \wfvs can be solved in polynomial time on co-comparability graphs~\cite[Theorem 3.7]{fvs_cocomparability}, we get the following from \cref{lem:fvs_same_time_as_wfvs,le:cocomp_p2}.

\begin{theorem}\label{thm:fvs_cocomparability}
    For a co-comparability graph $G$ there is an algorithm that computes a minimum feedback vertex set of $G \boxtimes P_\ell$ for any $\ell \in \mathbb{N}$ in polynomial time.
\end{theorem}

\subsection{Computational complexity for even path length}\label{se:fvs_even_length}

Our results so far are independent of the length of the path $P_\ell$ in $G \boxtimes P_\ell$.
Next, we show that if $\ell$ is even, \fvs on $G\boxtimes P_\ell$ is equivalent to \vc on $G$.

\begin{lemma}\label{le:fvs_vertex_cover_even}
    Let $\tau$ be the size of a minimum vertex cover of a graph $G$ and let $f$ be the size of a minimum feedback vertex set of $G \boxtimes P_{2\ell}$.
    Then, it holds that $2\ell\tau = f$.
\end{lemma}
\begin{proof}
    Let $P_{2\ell} = (p_1,\ldots,p_{2\ell})$.
    For any vertex cover $A$ of $G$ consider the set $F = \{ (v,p_i) \mid v \in A, i \in [2\ell]\}$.
    Then $(G \boxtimes P_{2\ell}) - F$ is a vertex-disjoint union of paths.
    This gives $f \leq 2\ell\tau$.

    Consider now a minimum feedback vertex set $F$ of $G\boxtimes P_2$.
    Since $\ceil{\frac{2\ell}{2}} = \floor{\frac{2\ell}{2}} = \ell$ , by \cref{obs:weighted_fvs}, we get that a minimum feedback vertex set of $G \boxtimes P_{2\ell}$ has size $\ell\cdot|F|$.
    Thus, it suffices to show that $|F| \geq 2 \cdot \tau$ where $\tau$ is the size of a minimum vertex cover of $G \boxtimes P_2$.
    
    To this end, we choose $F$ minimizing the size of $\project{F}{G}$, i.e., maximizing the number of vertices $v \in V(G)$ with $v \notin \project{F}{G}$.
    Let $T = (G \boxtimes P_2) - F$ be forest obtained by removing $F$.
    Clearly, $T$ contains no isolated vertices, as every vertex in $G \boxtimes P_2$ has a twin.
    Consider a vertex $v$ that is a leaf in $T$ (i.e., a vertex of degree~$1$).
    Let $w$ be the neighbor of $v$ in $T$.
    
    First assume that $\project{w}{G} \neq v.G$.
    Let $v'$ be the other copy of $v.G$ in $G \boxtimes P_2$.
    As $T$ has no triangles, $v'$ is not in $T$, i.e., $v' \in F$.
    Now observe that $F' = F - v' + w$ is again a feedback vertex set of $G \boxtimes P_2$.
    As $|\project{F'}{G}| < |\project{F}{G}|$, this contradicts the choice of $F$.

    Thus we have $w.G = v.G$, i.e., the only neighbor of $v$ in $T$ is the other copy of $v.G$ in $G \boxtimes P_2$.
    As $v$ and $w$ are twins in $G \boxtimes P_2$, and $T$ contains no triangles, $v$ and $w$ have no other neighbor in $T$, i.e., they form a component of size~$2$ in $T$.
    As this holds for every leaf in $T$, we get that $T$ is a matching, with all edges going between the two copies of $G$.
    In particular, restricting $F$ to the vertices of either copy yields a vertex cover of $G$, giving us that $|F| \geq 2\tau$, i.e., that $|F|$ is at least twice the size of a vertex cover of $G$.      
\end{proof}

\begin{corollary}\label{cor:vertex_cover}
    For every $\ell \in \mathbb{N}$ and every graph class $\mathcal{G}$, \fvs on $\{G \boxtimes P_{2\ell} \mid G \in \mathcal{G}\}$ is \NP-hard if and only if \vc on $\mathcal{G}$ is \NP-hard.
\end{corollary}

\cref{cor:vertex_cover} gives us graph classes $\mathcal{G}$ such that \fvs is in \P on $\mathcal{G}$ and \NP-hard on $\{G \boxtimes P_{2\ell} \mid G \in \mathcal{G}\}$ for every $\ell \in \mathbb{N}$; for example for the class $\mathcal{G}$ of all graphs with maximum degree~$3$~\cite{fvs_max_degree_3}.
Conversely, there are graph classes $\mathcal{G}$ such that \fvs is \NP-hard on $\mathcal{G}$ and in \P on $\{G \boxtimes P_{2\ell} \mid G \in \mathcal{G}\}$ for every $\ell \in \mathbb{N}$; for example for the class $\mathcal{G}$ of all bipartite graphs~\cite{fvs_bipartite}.

\subsection{Computational complexity for odd path length}\label{se:fvs_odd_length}

Here we consider \fvs on $G \boxtimes P_\ell$ when $\ell$ is odd.
In this case, it is useful to take a slightly different perspective.
Recall that by \cref{le:fvs_alternating} there is a minimum feedback vertex set $F$ of $G \boxtimes P_\ell$ that is alternating and such that $\project{(F \cap V_1)}{G} \subseteq \project{(F \cap V_2)}{G}$.
Let $F$ be such minimum feedback vertex set.
Then, for each vertex $v \in V(G)$, there are either $0,\floor{\frac{\ell}{2}},$ or $\ell$ copies of $v$ in $F$.
Equivalently, for each vertex $v \in V(G)$, there are either $0,\ceil{\frac{\ell}{2}},$ or $\ell$ copies of $v$ in the forest $T = (G \boxtimes P_\ell) - F$.
Consider a vertex $v \in V(G)$ such that all $\ell$ copies of $v$ are in $T$.
Then, as $T$ contains no triangles, no copy of any neighbor of $v$ is in $T$.
Thus, every vertex of degree~$0$ in $\project{T}{G}$ has $\ell$ copies in $T$, while every other vertex of $G$ has $\ceil{\frac{\ell}{2}}$ or $0$ copies in $T$.
That is, we can rephrase the problem in the following way.
For a forest $T$ and an integer $\ell \in \mathbb{N}$, we define the \emph{$\ell$-weight $w_\ell(T)$} as $\ceil{\frac{\ell}{2}}\cdot|V(T)| + \floor{\frac{\ell}{2}}\cdot i(T)$ where $i(T)$ is the number of isolated vertices of $T$. 
We define the corresponding decision problem as follows.
For an integer $\ell$, the \emph{$\ell$-weighted \fvs problem} asks for a given graph $G = (V,E)$ and integer $k$, whether there exists a set of vertices $U \subseteq V$ such that $G-U$ is a forest and $w_\ell(G-U) \geq k$.
Then, we can make the following observation.

\begin{observation}\label{obs:ell_weighted_forest}
    For every graph $G$ and every $\ell\in\mathbb{N}$, the maximum $\ell$-weight of an induced forest in $G$ is the same as the size of a maximum induced forest in $G \boxtimes P_\ell$.
\end{observation}

\begin{figure}
    \centering
    \includegraphics{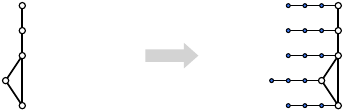}
    \caption{On the left a graph $G$.
    On the right the $3$-hair extension of $G$.}
    \label{fig:l_hair_extension}
\end{figure}

For a graph $G$ and an integer $k \in \mathbb{N}$ we define the \emph{$k$-hair extension} $H$ of $G$ as the graph that is obtained from $G$ by attaching to each vertex $v$ of $G$ a path $(h_1,\dots,h_k)$, i.e., each vertex $v \in V(G)$ has one additional neighbor in $H$ and each vertex $v$ in $H-G$ has either one or two neighbors in $H$.
An example for this is depicted in \cref{fig:l_hair_extension}.

\begin{lemma}\label{le:hair_extension}
    Let $\ell > 1$ be odd, $G$ be a graph, and $H$ be the $(2\ell-1)$-hair extension of $G$.
    Then $k$ is the size of a maximum induced forest in $G$, if and only the maximum $\ell$-weight of an induced forests of $H$ is $\frac{\ell+1}{2}\cdot (k + (2\ell-1) \cdot |V(G)|)$.
\end{lemma}
\begin{proof}
    Consider an induced forest $F$ in $H$ of maximum $\ell$-weight that maximizes the number of vertices that are in $H-G$.
    We claim that $V(H-G) \subseteq V(F)$, i.e., $F$ contains all paths that are attached to $G$.
    To see this, consider the path $(h_1,\dots,h_{2\ell-1})$ that is attached to some vertex $v\in V(G)$.
    Let $d$ be the number of vertices in $h_1,\dots,h_{2\ell-1}$ that are not in $F$.
    If $d=0$, we are done.
    If $d \neq \ell$ at most $d$ of $h_1,\dots,h_{2\ell-1}$ and $v$ are isolated in $F$.
    Thus, the total weight contributed by the path (and the potential extra weight due to $v$ being isolated) is $(2\ell - 1 - d)\frac{\ell+1}{2} + d\frac{\ell-1}{2}$ which is strictly smaller than $(2\ell-1)\frac{\ell+1}{2}$, which we obtain by adding all of the path to $F$.
    Assume now that $d = \ell$.
    Then, at most $d+1$ of $h_1,\dots,h_{2\ell-1}$ are isolated in $F$.
    Thus, the total weight contributed by the path (and the potential extra weight due to $v$ being isolated) is $(2\ell - 1 - \ell)\frac{\ell + 1}{2} + (\ell+1)\frac{\ell-1}{2}= (2\ell-2)\frac{\ell+1}{2}$ which is once again smaller than $(2\ell-1)\frac{\ell+1}{2}$.
    That is, we can add the entire path to $F$ and obtain the same weight.
    Therefore, we may assume in the following that indeed $V(H-G) \subseteq V(F)$.
    In particular this means that $F$ contains no isolated vertices.

    Clearly, if $G \cap F$ is not a maximum induced forest of $G$, taking the union of a larger induced forest in $G$ and all attached paths yields a larger $\ell$-weighted induced forest in $H$.
    Thus, the graph $G \cap F$ is a maximum induced forest in $G$ and we can partition $F$ into a maximum induced forest of $G$ and the $|V(G)|$ attached paths with $2\ell-1$ vertices each.
    This gives us that $w_\ell(F)= \frac{\ell+1}{2}\cdot (k + (2\ell-1) \cdot |V(G)|)$ as desired.
\end{proof}

From \cref{le:hair_extension} we almost immediately get the following theorem.

\begin{theorem}\label{thm:fvs_hardness_odd}
    Let $\mathcal{G}$ be graph class such that for every $G \in \mathcal{G}$ and every $k\in\mathbb{N}$ the $k$-hair extension $H$ of $G$ is also in $\mathcal{G}$, and such that \fvs is \NP-hard on $\mathcal{G}$.
    Then, for every $\ell \in \mathbb{N}$ \fvs is \NP-hard on the class $\{ G \boxtimes P_{2\ell-1} \mid G \in \mathcal{G}\}$.
\end{theorem}
\begin{proof}
    Let $\ell > 1$ be odd.
    Consider an instance of the \fvs problem consisting of a graph $G\in \mathcal{G}$ and integer $k\in \mathbb{N}$.
    Let $H$ be the $(2\ell-1)$-hair extension of $G$.
    Then, by \cref{le:hair_extension}, $G$ has an induced forest of size at least $k$ if and only if $H$ has an induced forest $F$ with $w_{\ell}(F)\geq \frac{\ell+1}{2}\cdot (k + (2\ell-1) \cdot |V(G)|)$.
    By \cref{obs:ell_weighted_forest}, this is equivalent to $H \boxtimes P_\ell$ having an induced forest of size at least $\frac{\ell+1}{2}\cdot (k + (2\ell-1) \cdot |V(G)|)$.
    That is, $(G,k)$ is a yes-instance of \fvs if and only if $(H \boxtimes P_\ell,\frac{\ell+1}{2}\cdot (k + (2\ell-1) \cdot |V(G)|))$ is a yes-instance of \fvs.
\end{proof}

\cref{thm:fvs_hardness_odd} in particular implies that for each $\ell \in \mathbb{N}$ that \fvs is \NP-hard on the class $\{ G \boxtimes P_{2\ell-1} \mid G \in \mathcal{G}\}$ if $\mathcal{G}$ is the class of all planar bipartite graphs~\cite{fvs_planar_bipartite}.
In particular, there are graph classes $\mathcal{G}$ such that \fvs is \NP-hard on $\{ G \boxtimes P_\ell \mid G \in \mathcal{G}\}$ for every odd $\ell \in \mathbb{N}$ but in \P on $\{ G \boxtimes P_\ell \mid G \in \mathcal{G}\}$ for every even $\ell \in \mathbb{N}$.

\section{Embedding Subdivisions of Subcubic Graphs into a Grid}
\label{sec:grid-embeddings}

In \cref{sec:max-cut,sec:edge-coloring} we show \NP-hardness for \mc and \ec on (induced) subgraphs of $P \boxtimes P$, respectively.
In both cases we reduce from the respective problem variant for subcubic graphs, i.e., graphs with maximum degree at most~$3$.
Starting with a given subcubic graph $G$, we embed a subdivision of $G$ into the strong grid $P \boxtimes P$.

\begin{definition}
    Let $G = (V,E)$ be a graph.
    A \emph{subdivision} of $G$ is a graph $G'$ that is obtained from $G$ by replacing each edge $e = uv$ of $G$ by an (induced) path $P(e)$ between $u$ and $v$, all of whose internal vertices are private to $P(e)$, i.e., disjoint from $P(e')$ for every edge $e' \neq e$ in $G$.
    We call $G'$ an \emph{even} subdivision, if for each edge $e$ of $G$, the path $P(e)$ contains an even number of vertices, i.e., $e$ is subdivided an even number of times (possibly 0).
\end{definition}

Every subcubic graph $G$ admits an orthogonal drawing on a polynomial grid, i.e., in which each edge is drawn as a polyline alternating between horizontal and vertical segments.
It is straightforward to convert an orthogonal drawing of $G$ to a subdivision $G'$ of $G$ that is an induced subgraph of $P \boxtimes P$.
As it is convenient later, let us make this more explicit here.

\begin{observation}
\label{obs:prel:planar-embedding}
    Let $G$ be a subcubic planar graph on $n$ vertices.
    Then there exists a (planar) subdivision $G'$ of $G$ that is an induced subgraph of $P_{2n} \Box P_{2n}$.
    In particular, $G'$ is an induced subgraph of $P_{4n} \boxtimes P_{4n}$ consisting only of diagonal edges.
\end{observation}

\begin{figure}
    \centering
    \includegraphics{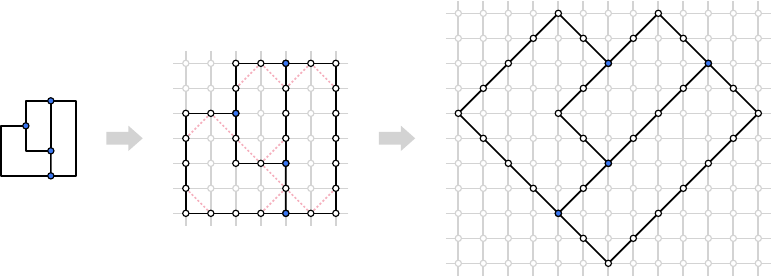}
    \caption{Embedding a subdivision $G'$ of a subcubic planar graph $G$ as a subgraph and induced subgraph of $P \boxtimes P$ according to \cref{obs:prel:planar-embedding}.
    On the left is the graph $G$.
    In the middle is the subdivision $G'$ as an induced subgraph of $P \Box P$.
    Note that it is not an induced subgraph of $P \boxtimes P$, indicated by the dotted edges.
    On the right is $G'$ as an induced subgraph of $P \boxtimes P$.}
    \label{fig:rotate-drawing}
\end{figure}

A subdivision $G'$ as claimed by \cref{obs:prel:planar-embedding} can for example be obtained as follows.
First, use the layout algorithm of Biedl and Kant~\cite{biedl_kant} to produce an orthogonal plane drawing $\Gamma$ of $G$ of size $n \times n$ which has vertices and edge bends on integer coordinates.
By placing subdivision vertices at integer coordinates for each edge, we obtain a subdivision $G_1$ of $G$ that is a subgraph of $P_n \Box P_n$.
Note that we can interchangeable talk about orthogonal drawings of $G$ and subdivisions of $G$ that are subgraphs of $P \Box P$.
Next, we scale the drawing $\Gamma$ by a factor of~$2$ to obtain a subdivision $G'$ of $G$ that is an induced subgraph of $P_{2n} \Box P_{2n}$.
In fact, $G'$ is obtained from $G_1$ by subdividing every edge of $G_1$ exactly once.
Finally, \cref{obs:grid-to-induced-strong-grid} gives that $G'$ is an induced subgraph of $P_{4n} \boxtimes P_{4n}$ consisting only of diagonal edges.
For an example refer to \cref{fig:rotate-drawing}.

\begin{observation}
\label{obs:prel:non-planar-embedding}
    Let $G$ be a subcubic (not necessarily planar) graph on $n$ vertices.
    Then there exists an even subdivision $G'$ of $G$ that is a subgraph of $P_{20n} \boxtimes P_{20n}$ with the following properties:
    \begin{itemize}
        \item At every original vertex of $G$ all incident edges are diagonal and uncrossed.
        \item At every vertex $v$ of $G'$ at most one incident edge at $v$ is crossed.
        \item Every orthogonal edge of $G'$ is neither adjacent to another orthogonal edge nor to a crossed edge.
        \item Every pair of crossing edges is an induced matching (of size 2) in $G'$.
        \item The induced subgraph $\tilde{G}$ of $P_{20n} \boxtimes P_{20n}$ on vertex-set $V(\tilde{G}) = V(G')$ contains all edges of $G'$ plus for each pair of crossing edges of $G'$ the remaining four edges which together form a $K_4$-subgraph in $\tilde{G}$.
    \end{itemize}
\end{observation}

\begin{figure}
    \centering
    \includegraphics{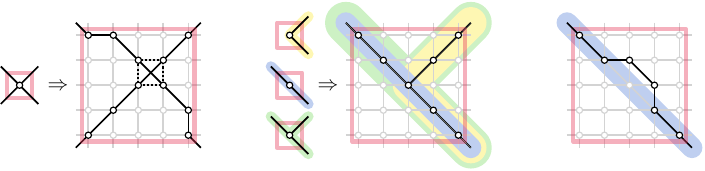}
    \caption{
        Embedding a subdivision $G'$ of a subcubic (not necessarily planar) graph $G$ as a subgraph of $P \boxtimes P$ according to \cref{obs:prel:non-planar-embedding}.
        Subdivision and original vertices (middle), crossings (left), and detours (right).
    }
    \label{fig:crossing-embedding}
\end{figure}

To obtain the subdivision $G'$ of $G$ as claimed in \cref{obs:prel:non-planar-embedding}, we modify the previous construction (for \cref{obs:prel:planar-embedding}) slightly:
We start with an orthogonal drawing of the subcubic graph $G$ in the plane with edge crossings at integer coordinates, subdivide, scale, and rotate as before.
The result can be seen as a subdivision $G_1$ of $G$ with an embedding $\Gamma_1$ into $P_{4n} \boxtimes P_{4n}$ with the following property.
For every pair $e_1,e_2$ of crossing edges in $G$ there are internal vertices $v_1$ in $P(e_1)$ and $v_2$ in $P(e_2)$ that are put in $\Gamma_1$ on the same grid point of $P_{4n} \boxtimes P_{4n}$.
To resolve this, we scale the drawing $\Gamma_1$ by a factor of~$5$ and add a detour for one of the two edges, as shown in left of \cref{fig:crossing-embedding}.
Finally, for every subdivision path $P(e)$ that contains an odd number of subdivision vertices, we add for exactly one of the subdivision vertices in $G_1$ a small detour as shows on the right of \cref{fig:crossing-embedding}.
The resulting graph $G'$ is an even subdivision of $G$, it is a subgraph of $P_{20n} \boxtimes P_{20n}$, and enjoys all desired properties.

\section{Max Cut}
\label{sec:max-cut}

Let $G = (V,E)$ be a graph and $C = (S,T)$ a bipartition of the vertex set of $G$.
Then we say that an edge $uv \in E$ with $u \in S, v \in T$ or $u \in T, v \in S$ is \emph{cut} by $C$.
Each edge not cut by $C$ is \emph{uncut}.
For a graph $G$ we denote by $mc(G)$ the maximum number of cut edges by any bipartition of $V$.
We denote by $\gamma(G)$ the minimum number of uncut edges by any bipartition of $V$.
Hence, for any graph $G$ we have $mc(G) = |E(G)| - \gamma(G)$.

The \mc problem asks for a given graph $G$ and an integer $k$, whether $mc(G) \geq k$.
The equivalent \minUncut problem (also known as Edge Bipartization) asks for a given graph $G$ and integer $k$ whether $\gamma(G) \leq k$.
Yannakakis~\cite{cubic_max_cut} shows that these problems remain \NP-hard when $G$ is restricted to be a cubic graph.
We use the \minUncut formulation in our proof that the problem remains \NP-hard when restricted to induced subgraphs of $P \boxtimes P$.
We first give a hardness proof for arbitrary subgraphs of $P \boxtimes P$, followed by outlining how to adapt the construction to work for induced subgraphs.
The overall plan is to embed a suitable subdivision $G'$ of a given graph $G$ into $P \boxtimes P$, as given in \cref{sec:grid-embeddings}.
For the reduction to work, we want that $G'$ and $G$ have the same uncut number.

\begin{lemma}
\label{lem:max-cut:subdivision}
    Let $G = (V,E)$ be a graph and let $G' = (V',E')$ be an even subdivision of $G$.
    Then $\gamma(G) = \gamma(G'),$ and there exists a bipartition of $G'$ attaining this value, which cuts every edge between two subdivision vertices.
\end{lemma}
\begin{proof}
    Let $C = (S,T)$ be a bipartition of $G$ that uncuts $k$ edges.
    We claim that then $G'$ has a bipartition $C' = (S', T')$ that uncuts $k$ edges.
    For this, we assign each vertex of $S$ to $S'$, each vertex of $T$ to $T'$ and for each subdivision path $P(uv)$ for $uv \in E(G)$, assign vertices alternatingly to the two parts, starting from $u$.
    Then each edge between two subdivision vertices is cut.
    Furthermore, since each $P(uv)$ has an even number of vertices, the edge incident to $v$ is uncut if and only if the edge $uv$ was uncut in $G$.
    This gives $\gamma(G') \leq \gamma(G)$.

    The reverse inequality is similar.
    Given a bipartition $C' = (S', T')$ of $G'$ that uncuts $k$ edges, we claim that the bipartition $C = (S' \cap V, T' \cap V)$ uncuts at most $k$ edges of $G$.
    For each edge $e$ uncut under $C$, both endpoints of the subdivision path $P(e)$ are contained in the same part of $C'$.
    Hence, by parity at least one edge of $P(e)$ is uncut under $C'$.
\end{proof}

Using \cref{lem:max-cut:subdivision}, the reduction from cubic graphs to subgraphs of $P \boxtimes P$ is simple:
Embed an even subdivision of the given graph into $P \boxtimes P$ to obtain a graph with the same uncut number.

\begin{theorem}\label{thm:max-cut-hardness-subgraphs}
    \mc restricted to subgraphs of $P \boxtimes P$ is \NP-hard.
\end{theorem}
\begin{proof}
    We reduce from \mc on cubic graphs.
    Let an instance $I_0 = (G, k_0)$ be given.
    We define the equivalent \minUncut instance $I := (G, k)$ with $k := |E(G)| - k_0$.
    By \cref{obs:prel:non-planar-embedding}, there exists an even subdivision $G'$ of $G$ that is a subgraph of $P_{20n} \boxtimes P_{20n}$.
    By \cref{lem:max-cut:subdivision}, the instances $I$ and $I' = (G', k)$ of \minUncut 
    are equivalent.
    Hence, the instance $I^* = (G', |E(G')| - k)$ of \mc
    is equivalent to $I_0 = (G,k_0)$.
\end{proof}

\begin{figure}[tb]
    \centering
    \includegraphics{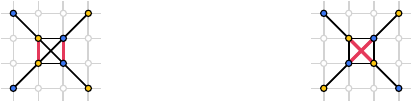}
    \caption{Left: Two cut edges crossing. Right: Two uncut edges crossing. In the induced setting, both uncut two edges in total each.
    The uncut edges are highlighted in red.}
    \label{fig:max-cut:induced-problem}
\end{figure}

To show that \mc is also \NP-hard for induced subgraphs of $P \boxtimes P$, we need to modify the subdivision $G'$ from the above argument.
For this, recall from \cref{obs:prel:non-planar-embedding} that the graph induced by $V(G')$ in $P \boxtimes P$ contains every edge of $G'$ plus for each crossing two additional edges to form a full $K_4$-subgraph instead.
This causes an issue as the $K_4$-subgraph admits a bipartition that ``hides'' edges originally uncut in $G'$, see \cref{fig:max-cut:induced-problem}.

We resolve this by a dedicated \emph{crossing gadget $G^\boxtimes$}, to be placed at each such crossing.
On its own, $\gamma(G^\boxtimes) = 8,$ i.e., each crossing gadget adds 8 uncut edges.
The gadget avoids the above issue by only attaining uncut number 8 when the corresponding original crossing edges were both cut.
Under a bipartition that uncuts one or two of the crossing edges, $G^\boxtimes$ contains at least 9 or 10 uncut edges respectively.
In this way, adding $c$ crossing gadgets increases the uncut number of the graph by exactly $8c$.

\begin{figure}[ht]
    \centering
    \includegraphics{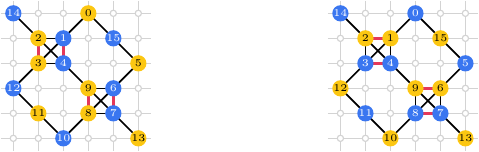}
    \caption{Mini-gadget $M$ with the two optimal bipartitions. The uncut edges are highlighted red.}
    \label{fig:max-cut:m-gadget}
\end{figure}

\begin{figure}[th]
    \centering
    \includegraphics{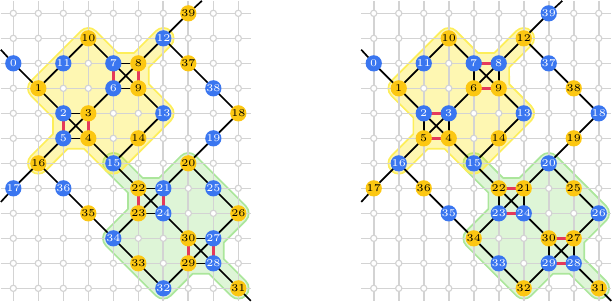}
    \caption{
        Crossing gadget $G^\boxtimes$ with two optimal cuts highlighted. 
        The uncut edges are highlighted in red.
        The two edge-disjoint copies of the gadget $M$ are highlighted in yellow and green, respectively.
    }
    \label{fig:max-cut:crossing}
\end{figure}

It remains to define and analyze $G^\boxtimes$.
The gadget is shown in \cref{fig:max-cut:crossing} and contains two copies of a smaller building block, the \emph{mini-gadget $M$}, shown in \cref{fig:max-cut:m-gadget}.
We analyze $G^\boxtimes$ bottom-up starting from its small parts.

\begin{observation}
\label{obs:max-cut:k4}
    It holds that $\gamma(K_4) = 2,$ attained exactly by the three bipartitions that contain two vertices in each part.
\end{observation}

\begin{lemma}
\label{lem:max-cut:m-gadget}
    It holds that $\gamma(M) = 4,$ attained exactly by the two bipartitions shown in \cref{fig:max-cut:m-gadget}.
\end{lemma}
\begin{proof}
    Note that $\gamma(M) \leq 4$ is given by \cref{fig:max-cut:m-gadget}, while $\gamma(M) \geq 4$ is given by \cref{obs:max-cut:k4}.

    To see that there are no other bipartitions certifying $\gamma(M) \leq 4$, note first that any bipartition of $M$ that uncuts four edges, uncuts no edge outside the two $K_4$'s.
    In particular, the vertices of the paths $1-6$ and $3-8$ are assigned alternatingly.
    As a result, when the top left $K_4$ is partitioned as in \cref{fig:max-cut:m-gadget}, the remaining vertex assignments follow.

    The only remaining case is that of $1,3$ being in the same set $V_i$.
    In this case, $6, 8$ are also in $V_i$ and so $2, 4, 7, 9$ are in $V_{1 - i}$.
    But then the edge $4, 9$ is a fifth uncut edge.
\end{proof}

We now review the full crossing gadget $G^\boxtimes$, shown in \cref{fig:max-cut:crossing}.
Note that $G^\boxtimes$ contains two edge-disjoint copies of the mini gadget $M$, and thus, \cref{lem:max-cut:m-gadget} gives $\gamma(G^\boxtimes) \geq 8$.

\begin{lemma}
\label{lem:max-cut:full-gadget-good-case}
    It holds that $\gamma(G^\boxtimes) = 8,$ attained exactly by the two bipartitions shown in \cref{fig:max-cut:crossing}.
\end{lemma}
\begin{proof}
    Let $C$ be a bipartition of $G^\boxtimes$ that uncuts at most eight edges.
    By \cref{lem:max-cut:m-gadget}, each copy of $M$ contains exactly four uncut edges and, thus, is assigned as shown in \cref{fig:max-cut:m-gadget}.
    Furthermore, no edge outside of the copies of $M$ is uncut.
    Fixing one of the two possible assignments for the top left copy of $M$, the remaining assignment follows by inspection.
\end{proof}

\begin{corollary}
    \label{cor:diagonal-not-uncut}
    Let $C = (V_0, V_1)$ be a bipartition of $G^\boxtimes$ that uncuts at most eight edges.
    Then $C$ uncuts none of the edges of the two diagonal paths $0-31$ and $17-39$.
\end{corollary}

By \cref{cor:diagonal-not-uncut}, the crossing gadget $G^\boxtimes$ attains uncut number 8 only when the original crossing edges are both cut.
Otherwise, at least nine edges of $G^\boxtimes$ are uncut.
It remains to show that when both original edges are uncut, then the uncut number of $G^\boxtimes$ rises to 10 to account for it.

\begin{lemma}
\label{lem:max-cut:10-edges}
    Let $C = (V_0, V_1)$ be a bipartition of $G^\boxtimes$ with $\{17, 39\} \subset V_i$ and $\{0, 31\} \subset V_j$ (possibly $i = j$).
    Then $C$ uncuts at least 10 edges.
\end{lemma}
\begin{proof}
    Assume for contradiction that $C$ uncuts at most nine edges of $G^\boxtimes$.
    If one of vertices 12 and 16 is contained in $V_i,$ then $C$ uncuts the corresponding edge and, thus, uncuts exactly four edges in each copy of $M$.
    By \cref{lem:max-cut:m-gadget}, it follows that vertices 1 and 31 are in the same set (via vertex 15).
    Thus, $1 \in V_j$ and the edge $\{0,1\}$ is uncut, a contradiction.

    It follows that $\{12, 16\} \subset V_{1 - i}$.
    \cref{lem:max-cut:m-gadget} now implies that $C$ uncuts at least five edges of the top left copy of $M$.
    Thus, $C$ uncuts exactly four edges of the bottom right copy of $M$ and no edges outside of either copy of $M$.
    It follows by paths $\{16, 36, 35, 34\}$ and $\{12, 37, 38, 18, 19, 20\}$ that $\{34, 20\} \subset V_i$.
    Then, by \cref{lem:max-cut:m-gadget}, $C$ uncuts more than four edges of the bottom right copy of $M$, the final contradiction.
\end{proof}

We are now ready to prove that removing a crossing gadget reduces the uncut number of a graph by at least 8.
The converse fact that adding a crossing gadget increases the uncut number of a graph by at most 8, shows that either bound is tight and allows us to then deduce the main theorem.
Let $G^\times$ be the subgraph of $G^\boxtimes$ on all diagonal edges in $G^\boxtimes$.

\begin{restatable}{lemma}{maxcutlemma}\label{lem:max-cut:minus-8-edges}
    Let $C^\boxtimes$ be a bipartition of $G^\boxtimes$ that uncuts $k$ edges.
    Then there exists a bipartition $C'$ of $G^\times$ that uncuts at most $k - 8$ edges and assigns vertices 0, 17, 31 and 39 to the same sets as in $C^\boxtimes$.
\end{restatable}

\begin{proof}
    The bipartition $C'$ is the obvious one: starting at the already assigned vertices 0 and 17, vertices of either path are partitioned alternately.
    To see that this suffices, we consider the different uncut counts of $C^\boxtimes$.
    If it uncuts eight edges (the minimum possible), then by \cref{lem:max-cut:full-gadget-good-case}, 0 and 31, and 17 and 39, are in opposite parts and the above partition of $G^\times$ uncuts no edges.
    If $C^\boxtimes$ uncuts nice edges, then by \cref{lem:max-cut:10-edges}, at least one of $(0,31)$ or $(17,39)$ are in opposite parts.
    Thus, $C'$ uncuts at most one edge.
    Finally, regardless of $C^\boxtimes$, $C'$ uncuts at most two edges which covers the case where $C^\boxtimes$ uncuts at least ten edges.
\end{proof}

With \cref{lem:max-cut:minus-8-edges} we can complete the hardness reduction for induced subgraphs of $P \boxtimes P$.

\begin{theorem}\label{thm:max-cut-hardness-induced}
    \mc on induced subgraphs of $P \boxtimes P$ is \NP-complete.
\end{theorem}
\begin{proof}
    The problem is clearly contained in \NP.
    To show \NP-hardness, we reduce from \mc on cubic graphs.
    Let an instance $I_0 = (G,k_0)$ be given.
    We transform it into the equivalent \minUncut instance $I = (G, k)$ with $k = |E(G_0)| - k_0$.
    We then transform $I$ into an instance $I' = (G', k')$ of \minUncut with $G'$ being an induced subgraph of $P \boxtimes P$.

    \begin{figure}
        \centering
        \includegraphics{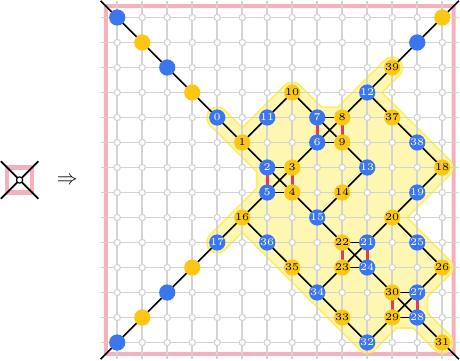}
        \caption{The gadget $G^\boxtimes$ (yellow) extended to be a subgraph of $P_{14} \boxtimes P_{14}$.}
        \label{fig:big-gadget-extended}
    \end{figure}
    
    For this, define $G^*$ as an even subdivision of $G$ that is a subgraph of $P_{20n} \boxtimes P_{20n}$, as given by \cref{obs:prel:non-planar-embedding}.
    We scale the entire drawing by a factor of~$7$, replacing each crossing of size $2 \times 2$ by a copy of the crossing gadget $G^\boxtimes$ inside a $14 \times 14$ area, as shown in \cref{fig:big-gadget-extended}, resulting in a graph $G' \subseteq P_{140n} \boxtimes P_{140n}$.
    
    Let $c$ be the number of crossings in $G^*$, and set $k' := k + 8 \cdot c$.
    We claim that $I' = (G', k')$ is a yes-instance of \minUncut if and only if $I = (G,k)$ is a yes-instance of \minUncut.

    Let $C$ be a bipartition of $V(G)$ that uncuts at most $k$ edges.
    By \cref{lem:max-cut:subdivision} there is a bipartition of $V(G^*)$ that uncuts at most $k$ edges and uncuts no edges between subdivision vertices.
    This bipartition can be extended to one of $V(G')$ by assigning the vertices of the various crossing gadgets according to \cref{fig:max-cut:crossing}.
    The resulting bipartition uncuts an extra eight edges per crossing gadget, so no more than $k + 8c$ in total.

    Now let $C' = (S',T')$ be a bipartition of $V(G')$ that uncuts at most $k + 8c$ edges.
    Then there exists a bipartition of $V(G^*)$ that uncuts at most $k + 8c - 8 \cdot c = k$ edges:
    Retain $C'$ outside of crossing gadgets, and for each of the $c$ crossing gadgets of $G'$, assign the corresponding crossing paths of $G^*$ according to \cref{lem:max-cut:minus-8-edges}.
    Then, by \cref{lem:max-cut:subdivision}, there exists a bipartition of $V(G)$ that uncuts at most $k$ edges.

    This shows that the \minUncut instances $I' = (G', k')$ and $I = (G,k)$ are equivalent, from which it follows that also \mc is \NP-hard on induced subgraphs of $P \boxtimes P$.
\end{proof}

\subsection{Max Cut on the Full Product}

In this subsection, we study \mc on the full product $G \boxtimes P_\ell$ for some graphs $G$ from some graph class $\mathcal{G}$.
We show that if $\mathcal{G}$ is the class of bipartite graphs, then \mc can be solved in polynomial time on $G \boxtimes P_\ell$ for every $G \in \mathcal{G}$ and every $\ell \in \mathbb{N}$.

As before, it is more convenient to consider the equivalent \minUncut problem.

\begin{theorem}
    \label{thm:main-mc-bipartite}
    Let $G$ be a connected bipartite graph on $n$ vertices and $m$ edges and let $P_\ell$ be the path on $\ell$ vertices.
    Then $\gamma(G \boxtimes P_\ell) = \min\{m \cdot \ell, n \cdot (\ell - 1)\}.$
    In particular, unless $G$ is a tree on at most $\ell$ vertices, it holds that $\gamma(G \boxtimes P_\ell) = n \cdot (\ell - 1).$
\end{theorem}

\begin{proof}
    We first show that $\gamma(G \boxtimes P_\ell) \leq \min\{m \cdot \ell, n \cdot (\ell - 1)\}$.
    To do this, let $P_\ell = (p_1,\dots,p_\ell)$ and denote the bipartition of $G$ by $(A,B)$, i.e., $V(G) = A \mathbin{\dot\cup} B$.
    Then, the cut obtained by taking every second copy of $G$ uncuts exactly the $m$ edges of $G$ inside each of the $\ell$ copies, showing that $\gamma(G \boxtimes P_\ell) \leq m \cdot \ell$.
    Such a partition is depicted in \cref{fig:bipartite_maxcut_upper_bound} on the left.
    Conversely, the cut obtained by taking every copy of $A$ uncuts for each of the $n$ vertices $v$ of $G$, each of the $\ell-1$ edges between two consecutive copies of $v$, showing that $\gamma(G \boxtimes P_\ell) \leq n \cdot (\ell - 1)$.
    Such a partition is depicted in \cref{fig:bipartite_maxcut_upper_bound} on the right.

    \begin{figure}
        \centering
        \includegraphics{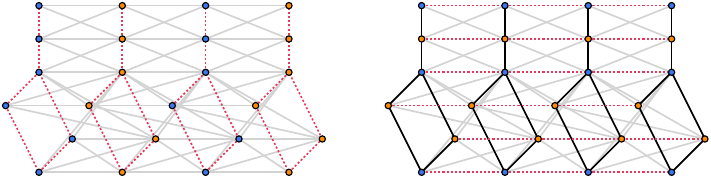}
        \caption{Two examples for a partition of the vertices of $G \boxtimes P_\ell$ for some bipartite graph $G$.
        The partition on the left uncuts $m \cdot \ell$ edges. 
        The partition of the right uncuts $n \cdot (\ell-1)$ edges.        
        Uncut edges are dotted and red.}
        \label{fig:bipartite_maxcut_upper_bound}
    \end{figure}

    It remains to show the lower bound $\gamma(G \boxtimes P_\ell) \geq \min\{m \cdot \ell, n \cdot (\ell - 1)\}$.
    For this, we find a set $\mathcal{S}$ of edge-disjoint triangles and $K_4$-subgraphs in $G \boxtimes P_\ell$.
    Since any bipartition uncuts at least one edge of each $K_3$ and at least two edges of each $K_4$, this yields a lower bound for the uncut number of $G \boxtimes P_\ell$.
    Our construction of $\mathcal{S}$ depends on whether $G$ is a tree or not.

    \smallskip
    
    \textbf{Case $G$ is not a tree:}
    We show that $\gamma(G \boxtimes P_\ell) \geq n \cdot (\ell-1)$.
    For this, let $G' \subseteq G$ be a spanning tree of $G$ plus a single edge.
    Orient the edges of the cycle of $G'$ in cyclic order, and orient every other edge towards the cycle.
    Similarly, orient $P_\ell$ from $p_1$ to $p_\ell$.
    Then $\mathcal{S} := \{\{(s, p_i), (s,p_{i + 1}), (t, p_i)\} \mid st \in G', p_ip_{i + 1} \in P_\ell \}$ is a set of $n \cdot (\ell - 1)$ edge-disjoint triangles in $G' \boxtimes P_\ell$.
    See \cref{fig:example_triangles_nontree} for an example.
    Under any bipartition of $V(G \boxtimes P_\ell)$ each triangle in $\mathcal{S}$ contains at least one uncut edge, and thus $\gamma(G \boxtimes P_\ell) \geq |\mathcal{S}| = n \cdot (\ell-1)$.

    \begin{figure}
        \centering
        \begin{subfigure}[t]{0.24\textwidth}
            \centering
            \includegraphics[page=1]{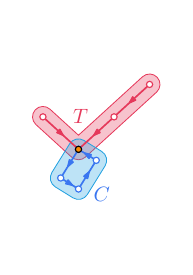}
            \subcaption{The graph $G$}
            \label{fig:example_triangles_nontree_G}
        \end{subfigure}%
        \begin{subfigure}[t]{0.25\textwidth}
            \centering
            \includegraphics[page=2]{figures/example_triangles.pdf}
            \subcaption{$G \Box P_3$}
            \label{fig:example_triangles_nontree_G_box_P}
        \end{subfigure}%
        \begin{subfigure}[t]{0.25\textwidth}
            \centering
            \includegraphics[page=3]{figures/example_triangles.pdf}
            \subcaption{$C \boxtimes P_3$}
            \label{fig:example_triangles_nontree_C}
        \end{subfigure}%
        \begin{subfigure}[t]{0.25\textwidth}
            \centering
            \includegraphics[page=4]{figures/example_triangles.pdf}
            \subcaption{$T \boxtimes P_3$}
            \label{fig:example_triangles_nontree_T}
        \end{subfigure}%
        \caption{In \subref{fig:example_triangles_nontree_G} a bipartite graph $G$ consisting of a cycle $C$ and a tree $T$, in \subref{fig:example_triangles_nontree_G_box_P} a depiction of $G \Box P_3$ with edges orientated according to the proof of \cref{thm:main-mc-bipartite}, in \subref{fig:example_triangles_nontree_C} the edge-disjoint triangles found in $C \boxtimes P_3$, and in \subref{fig:example_triangles_nontree_T} the edge-disjoint triangles found in $T \boxtimes P_3$.}
        \label{fig:example_triangles_nontree}
    \end{figure}

    \smallskip
    
    \textbf{Case $G$ is a tree:}
    Let $x := \min\{m, \ell - 1\}$ be the number of edges of the smaller of the two factor graphs.
    Let $e_1,\ldots,e_m$ be the edges of $G$.
    Again, we define a set $\mathcal{S}$ of edge-disjoint $K_3$- and $K_4$-subgraphs in $G \boxtimes P_\ell$.
    For each $i \in [m]$, $j \in [\ell-1]$ define by $H_{i,j}$ the \emph{square} $e_i \boxtimes p_jp_{j+1}$ contained in $G \boxtimes P_\ell$.
    Note that two squares $H_{i,j}, H_{i',j'}$ share an edge if and only if $i = i'$ and $p_jp_{j+1}$ is adjacent to $p_{j'}p_{j'+1}$ in $P_\ell$, or $j = j'$ and $e_i$ is adjacent to $e_{i'}$ in $G$.
    
    Our set $\mathcal{S}$ contains the $k$ edge-disjoint squares $H_i := H_{i,i}$ for $i = 1,\ldots,x$.
    Additionally, for each square $H_{i,j}$ with $i \neq j$, we shall add to $\mathcal{S}$ a triangle obtained from $H_{i,j}$ by removing one of the four vertices.
    Thus, $\mathcal{S}$ shall contain $x$ squares and $m \cdot (\ell-1) - x$ triangles.
    If all squares and triangles in $\mathcal{S}$ are edge-disjoint, it follows that $\gamma(G \boxtimes P_\ell) \geq 2x + m \cdot (\ell-1) - x = m\cdot(\ell-1) + x = \min\{m \cdot \ell, n \cdot (\ell - 1)\}$, as desired.

    It remains to define a triangle $\Delta_{i,j} \subset H_{i,j}$ whenever $i \neq j$ such that $H_1,\ldots,H_x$ and these triangles are pairwise edge-disjoint.
    To this end, we root $P_\ell$ at $p_1$ and $G$ at some arbitrary leaf $r$.
    Consider now a square $H_{i,j}$, i.e., for $e_i = uv$ the $K_4$ induced by $\{u,v\}\times \{p_j,p_{j+1}\}$.
    For now, assume that $i,j \leq k$, i.e., $H_i,H_j \in \mathcal{S}$.
    Then, either $(u,p_j)$ and $(v,p_j)$, or $(u,p_{j+1})$ and $(v,p_{j+1})$ are closer to $H_i$ in $G \boxtimes P_\ell$.
    Similarly, either $(u,p_j)$ and $(u,p_{j+1})$, or $(v,p_j)$ and $(v,p_{j+1})$ are closer to $H_j$ in $G \boxtimes P_\ell$.
    We obtain $\Delta_{i,j} \subset H_{i,j}$ by removing from $H_{i,j}$ the vertex that is closer in both cases.
    If $H_i$ does not exist, i.e., $i > x$, we instead take the distance to $e_i \boxtimes \{p_1\}$ in $G \boxtimes P_\ell$.
    Analogously, if $H_j$ does not exist, i.e., $j > x$, we instead take the distance to $\{r\} \boxtimes p_jp_{j+1}$ in $G \boxtimes P_\ell$.
    See \cref{fig:example_triangles_tree_T} for an example.

    \begin{figure}
        \centering
        \includegraphics{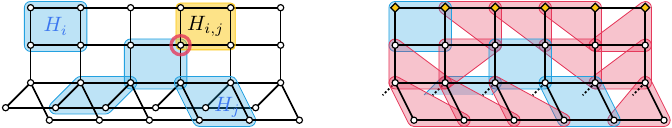}
        \caption{An example for a graph $G \boxtimes P_4$ where $G$ is a tree.
        For the sake of visual clarity the diagonal edged are omitted.
        On the left the chosen copies of $K_4$ are highlighted in blue.
        For the square $H_{i,j}$ (yellow) we delete the vertex on the bottom-left, as it is closest to $H_i$ and $H_j$.
        The final packing of triangles and copies of $K_4$ is depicted on the right.
        The chosen leaf $r$ in order to obtain the triangles in the rightmost copy of $G$ is the one corresponding to the orange diamonds.
        In this example, this only effects the rightmost triangles.
        For the sake of visual clarity some vertices and edges are omitted.}
        \label{fig:example_triangles_tree_T}
    \end{figure}

    It is left to argue that the triangles $\{\Delta_{i,j} \mid i \neq j\}$ and squares $\{H_{i,j} \mid i = j\}$ are pairwise edge-disjoint.
    The squares are edge-disjoint since we have at most one square for each edge of $G$ and at most one square for each edge of $P_\ell$.
    Consider now a square $H_i$ and a triangle $\Delta_{j,k}$.
    Clearly, if $i \notin \{j,k\}$ we have that $H_i$ and $\Delta_{j,k}$ share at most one vertex.
    We assume that $i=j$ (the case $i=k$ is analogous).
    Then, the two vertices of $H_{j,k}$ incident to $H_i$ are closer to $H_i$ than the other two.
    Thus, one of them was deleted to obtain $\Delta_{j,k}$, giving us that $H_i$ and $\Delta_{j,k}$ are edge-disjoint.

    Finally, consider two triangles $\Delta_{i,j}$ and $\Delta_{i',j'}$.
    Clearly, if $i \neq i'$ and $j \neq j'$ we have that $\Delta_{i,j}$ and $\Delta_{i',j'}$ share at most one vertex.
    We assume that $i=i'$ (the case $j=j'$ is analogous).
    If $p_jp_{j+1}$ and $p_{j'}p_{j'+1}$ are independent, then $\Delta_{i,j}$ and $\Delta_{i',j'}$ are vertex-disjoint.
    Thus, assume that $p_jp_{j+1}$ and $p_{j'}p_{j'+1}$ share a vertex $p$ in $P_\ell$. 
    This means that $H_{i,j}$ and $H_{i,j'}$ share $e_i \boxtimes \{p\}$ in $G \boxtimes P_\ell$. 
    Now let $A = H_i$ if $i \leq x$ and $A = e_i \boxtimes \{p_1\}$ if $i > x$.
    Assume without loss of generality that $A$ is at least as close to $H_{i,j'}$ as to $H_{i,j}$.
    Then, the two vertices in $H_{i,j}$ that correspond to $p$ are closer to $A$ than the two vertices corresponding to the other endpoint of $p_jp_{j+1}$.
    Thus, one of the two shared vertices of $H_{i,j}$ and $H_{i,j'}$ is removed to obtain $\Delta_{i,j}$, and $\Delta_{i,j}$ and $\Delta_{i,j'}$ are edge-disjoint.
\end{proof}

\begin{corollary}\label{cor:maxcut_bipartite}
    There is an algorithm that computes for any $\ell \in \mathbb{N}$ and any bipartite graph $G$ a maximum cut of $G \boxtimes P_\ell$ in linear time.
\end{corollary}

\begin{figure}
    \centering
    \includegraphics{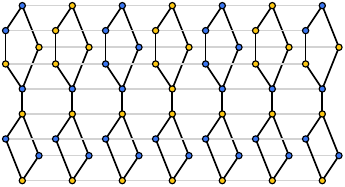}
    \caption{An example for a graph $G\boxtimes P_{7}$ with a partition of the vertices corresponding to a maximum cut.
    Every alternating partition of the vertices results in a smaller cut.
    For the sake of visual clarity, we depict $G \Box P_{7}$ instead of $G \boxtimes P_{7}$.}
    \label{fig:counterexample_alternating_maxcut}
\end{figure}

\cref{thm:main-mc-bipartite} additionally gives that if $G$ is bipartite, then $G \boxtimes P_\ell$ admits a maximum cut with a bipartition $C=(S,T)$ such that $S$ and $T$ are alternating, i.e., $S$ contains the same vertices of every odd copy of $G$ and the same vertices of every even copy of $G$.
We have an analogous statement for \fvs (\cref{le:fvs_alternating}).
For every graph $G$ and every $\ell \in \mathbb{N}$ the graph $G \boxtimes P_\ell$ admits an optimal feedback vertex set that is alternating.
For \mc, however, this statement is false in general (for not necessarily bipartite graphs $G$).
For example, \cref{fig:counterexample_alternating_maxcut} depicts the strong product $G \boxtimes P_7$ of a cactus graph $G$ and $P_7$, such that every alternating bipartition of $G \boxtimes P_7$ results in a suboptimal cut.

\section{Edge Coloring}
\label{sec:edge-coloring}

Given a graph $G = (V,E)$, a \emph{proper $k$-edge-coloring} of $G$ is an assignment $c \colon E \to [k]$ such that every pair $e_1,e_2$ of edges in $G$ that shares a common vertex receives distinct colors, i.e., $c(e_1) \neq c(e_2)$.
The \emph{chromatic index} of $G$ (also called edge-chromatic number), denoted by $\chi'(G)$, is the smallest integer $k$ such that $G$ admits a proper $k$-edge-coloring.
Given a graph $G$, the \emph{\ec problem} asks whether $\chi'(G) \leq 3$.
If $\Delta(G)$ denotes the maximum degree of $G$, then (clearly) $\chi'(G) \geq \Delta(G)$.
On the other hand, Vizing's Theorem~\cite{vizings_theorem} states $\chi'(G) \leq \Delta(G) + 1$ for every graph $G$.
Hence, we may assume throughout that $\Delta(G) = 3$, in particular that $G$ is subcubic.

For our reductions we start with a subcubic graph $G$ and construct a subcubic graph $\hat{G}$ that is a subgraph of the strong grid $P \boxtimes P$ in such a way that $\chi'(\hat{G}) = \chi'(G)$.
Our construction heavily relies on the subdivisions $G'$ of $G$ given by \cref{obs:prel:planar-embedding,obs:prel:non-planar-embedding} in \cref{sec:grid-embeddings}, and indeed enjoy very similar properties.
If $G$ is planar, then $\hat{G}$ is an induced subgraph of $P \boxtimes P$.
And if $G$ is not necessarily planar, then $\hat{G} \subset P \boxtimes P$ is almost an induced subgraph with only the pairs of crossing edges inducing a $K_4$-subgraph instead of a $2K_2$-subgraph.

\begin{proposition}
    \label{prop:edge-coloring-gadget}
    Let $G$ be a subcubic graph on $n$ vertices.
    Then we can construct in polynomial time a graph $\hat{G}$ that is a subgraph of $P_{180n} \boxtimes P_{180n}$ such that $\chi'(\hat{G}) = \chi'(G)$.

    Moreover, if $G$ is planar, then $\hat{G}$ is planar and an induced subgraph of $P_{180n} \boxtimes P_{180n}$.
\end{proposition}
\begin{proof}
    We are given the graph $G = (V,E)$ with $n = |V|$.
    If $G$ is planar, let $G'$ be the (planar) subdivision of $G$ given by \cref{obs:prel:planar-embedding}, i.e., $G'$ is an induced subgraph of $P_{4n} \boxtimes P_{4n} \subset P_{20n} \boxtimes P_{20n}$ consisting only of diagonal edges.
    Otherwise (if $G$ is not necessarily planar), let $G' \subset P_{20n} \boxtimes P_{20n}$ be the subdivision of $G$ given by \cref{obs:prel:non-planar-embedding}.
    In any case, we scale the strong grid by a factor of~$9$, replacing each grid point in $P_{20n} \boxtimes P_{20n}$ by a $9 \times 9$ strong grid in $P_{180n} \boxtimes P_{180n}$.
    In doing so, we replace each vertex of $G'$ by a subgraph as shown in \cref{fig:3-edge-coloring-gadgets}.
    The result is the graph $\hat{G} \subset P_{180n} \boxtimes P_{180}$ with maximum degree $\Delta(\hat{G}) = 3$.
    Moreover, if $G$ is planar, then $\hat{G}$ is planar and an induced subgraph of $P_{180n} \boxtimes P_{180n}$.

    \begin{figure}
        \centering
        \includegraphics{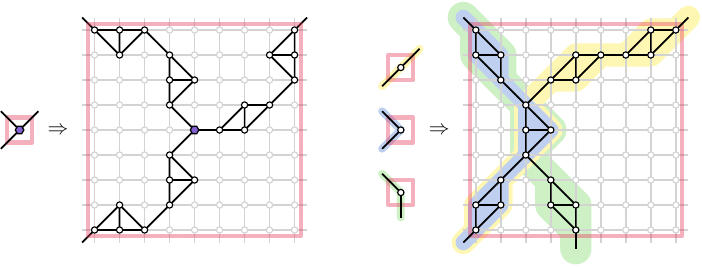}
        \caption{
            Obtaining $\hat{G}$ from a subdivision $G' \subset P \boxtimes P$ of $G$ by replacing each grid point by a $9 \times 9$ strong grid.
        }
        \label{fig:3-edge-coloring-gadgets}
    \end{figure}

    It remains to argue that $\chi'(\hat{G}) = \chi'(G)$.
    As $\Delta(\hat{G}) = 3 = \Delta(G)$, it is enough to show that $G$ admits a proper $3$-edge-coloring if and only if $\hat{G}$ admits a proper $3$-edge-coloring.
    
    For every edge $e = uv$ of $G$ there is a subgraph $Q(e)$ in $\hat{G}$ that is a sequence
    \[
        e_1,S_1,e_2,S_2,\ldots,e_{t-1},S_{t-1},e_t \quad \text{of edges } e_1,\ldots,e_t \text{ and } K^-_4\text{-subgraphs\footnote{$K^-_4$ is the graph obtained from $K_4$ by deleting one edge.} } S_1,\ldots,S_{t-1}
    \]
    for some integer $t \geq 2$.
    The first and last edges $e_1,e_t$ in $Q(e)$ are incident to the respective endpoints $u,v$ of edge $e$.
    Observe that $Q(e)$ admits a proper $3$-edge-coloring.
    Moreover, in every proper $3$-edge-coloring of $Q(e)$ the edges $e_1$ and $e_t$ receive the same color.
    Thus, if $G$ admits a proper $3$-edge-coloring, then so does $\hat{G}$.
    And if $\hat{G}$ admits a proper $3$-edge-coloring $\hat{c} \colon E(\hat{G}) \to [3]$, then we obtain a proper $3$-edge-coloring $c$ of $G$ by setting for every edge $e$ of $G$ its color to $c(e) = \hat{c}(e_1) = \hat{c}(e_t)$, where $e_1,e_t$ are the first and last edge in $Q(e)$.
\end{proof}

With \cref{prop:edge-coloring-gadget} at hand, we can easily derive the main result of this section.

\begin{theorem}
    \label{thm:edge-coloring-main-thm}
    Each of the following holds.
    \begin{itemize}
        \item \ec is \NP-complete for subgraphs of $P \boxtimes P$.
        \item \ec for induced subgraphs of $P \boxtimes P$ is polynomially equivalent to \ec for planar graphs.
    \end{itemize}    
\end{theorem}
\begin{proof}
    \ec is known to be \NP-complete for subcubic graphs.
    Given a subcubic graph $G$, we construct with \cref{prop:edge-coloring-gadget} a graph $\hat{G} \subseteq P \boxtimes P$ such that $\chi'(\hat{G}) = \chi'(G)$.
    This is a polynomial-time reduction and shows that \ec is \NP-hard for subgraphs of $P \boxtimes P$.
    (The problem is clearly in \NP.)

    In the same way, given a subcubic planar graph $G$, we obtain with \cref{prop:edge-coloring-gadget} in polynomial time an induced subgraph $\hat{G}$ of $P \boxtimes P$ with $\chi'(\hat{G}) = \chi'(G)$.
    Thus, \ec for planar graphs can be polynomially reduced to \ec for induced subgraphs of $P \boxtimes P$.
    For the other direction it is enough to observe that if $G$ is an induced subgraph of $P \boxtimes P$ with maximum degree $\Delta(G) \leq 3$, then $G$ is actually planar.
    Indeed, if $G$ contains two crossing (diagonal) edges $u_1v_1$ and $u_2v_2$, then $\{u_1,v_1,u_2,v_2\}$ induce a $K_4$-subgraph in $G$.
    As $G$ is subcubic, this $K_4$ is an entire connected component of $G$.
    As all connected components without crossing edges are planar by definition, $G$ is a planar graph.
    Thus the trivial reduction (simply keep the given graph $G$) shows that \ec for induced subgraphs of $P \boxtimes P$ reduces to \ec for planar graphs.
\end{proof}

\section{Conclusions}
\label{sec:conclusion}
\label{app:further_problems}

In this paper we considered the computational complexity of \fvs, \mc, and \ec on subgraphs of $G \boxtimes P_\ell$, induced subgraphs of $G \boxtimes P_\ell$, and the full graph $G \boxtimes P_\ell$.
We identified cases that are polynomial-time solvable and cases that are \NP-complete.
See again \cref{fig:overview} for an overview of our results.
A few cases remain open.

\begin{question}
    What is the computational complexity of \ec for induced subgraphs of the strong grid $P \boxtimes P$?
    By \cref{thm:edge-coloring-main-thm} this is polynomially equivalent to \ec for planar graphs.
\end{question}

\begin{question}
    What is the computational complexity of \mc for $G \boxtimes P_\ell$ if \mc is efficiently solvable for $G$?
    By \cref{thm:main-mc-bipartite} there is a linear-time algorithm for \mc on $G \boxtimes P_\ell$, if $G$ is bipartite.
\end{question}

\begin{question}
    What is the computational complexity of \fvs for $G \boxtimes P_\ell$ for odd $\ell$ if \fvs is efficiently solvable for $G$?
    By \cref{obs:ell_weighted_forest} this is polynomially equivalent to $\ell$-weighted \fvs for $G$.
\end{question}

To conclude, let us also briefly discuss some other classical \NP-complete graph problems.

\subparagraph*{Clique.}

Given a graph $G$ and some integer $k$, the \emph{\textsc{Clique} problem} asks, whether $G$ contains a clique on~$k$ vertices, i.e., a set of~$k$ pairwise adjacent vertices.
For subgraphs of $P \boxtimes P$ this can clearly be decided in time $O(n^4)$ as every clique in $P \boxtimes P$ contains at most four vertices.
More generally, if $H \subseteq G \boxtimes P_\ell$, then every clique in $H$ is contained in two consecutive copies of $G$.
This gives an algorithm for \textsc{Clique} in $H$ that is \FPT in the treewidth of $G$ as follows.
Consider every pair of consecutive copies of $G$ and the corresponding subgraph $H' \subseteq H$.
In particular, we have $H' \subseteq G \boxtimes P_2$ and, thus, the treewidth of $H'$ is at most twice the treewidth of $G$ plus one.%
\footnote{A more general statement is for example shown by Hickingbotham and Wood~\cite[Theorem~27]{treewidth_strong_product_p2}.}
Using a standard dynamic programming algorithm on a tree decomposition of $H'$ for each such graph $H'$ yields an \FPT algorithm for $H$.

\subparagraph*{Independent Set.}

Given a graph $G$ and some integer $k$, the \emph{\textsc{Independent Set} problem} asks, whether $G$ contains an independent set on~$k$ vertices, i.e., a set of~$k$ pairwise non-adajcent vertices.
\textsc{Independent Set} is known to be \NP-complete on induced subgraphs of $P \boxtimes P$~\cite{independent_set_induced_subgraphs}.
For the full product $G \boxtimes P_\ell$, one can observe that $G \boxtimes P_\ell$ admits a maximum independent set that consists of a maximum independent set of $G$ in every second copy of $G$.
This gives that \textsc{Independent Set} on $G \boxtimes P_\ell$ is equivalent to \textsc{Independent Set} on $G$.
For a more in depth argument refer for example to the Bachelor's thesis of Rapp~\cite[Theorem~3.2]{rapp_bachelor}.

\subparagraph*{Vertex Coloring.}

Given a graph $G = (V,E)$ and some integer $k$, a \emph{proper $k$-vertex-coloring} of $G$ is a coloring $c \colon V \to [k]$ such that for every edge $e = uv$ in $G$ it holds $c(u) \neq c(v)$.
The \emph{chromatic number} of $G$, denoted by $\chi(G)$, is the smallest integer $k$ such that $G$ admits a proper $k$-vertex-coloring, and 
the \emph{\textsc{Vertex Coloring} problem} asks whether $\chi(G) \leq k$.

For the full product $G \boxtimes P_\ell$, it holds%
\footnote{A more general statement is for example shown by Esperet and Wood~\cite[Lemma~9]{chromatic_number_strong_product_upper_bound}.}
that $\chi(G \boxtimes P_\ell) \leq 2 \cdot \chi(G)$, where equality holds if $G$ is bipartite (and contains at least one edge).
If $G$ is not bipartite, then $\chi(G \boxtimes P_\ell) \geq 5$~\cite[Theorem~1]{chromatic_number_strong_product_lower_bound}.
For every $\ell \geq 2$ it follows that $\chi(G \boxtimes P_\ell) \leq 4$ if and only if $G$ is bipartite.
In particular, \textsc{Vertex Coloring} is trivial on $G \boxtimes P_\ell$ for $k \leq 4$ or if $G$ is bipartite.
For $k \geq 5$ and non-bipartite $G$ the \textsc{Vertex Coloring} problem appears interesting.
For example, the bound $\chi(G \boxtimes P_\ell) \leq 2 \cdot \chi(G)$ is not tight for $G = C_5$.

On subgraphs of $P \boxtimes P$, \textsc{Vertex Coloring} is trivial for $k > 3$, since $\chi(P \boxtimes P) = 4$.
For $k=3$, \textsc{Vertex Coloring} is \NP-complete even on induced subgraphs of $P \boxtimes P$~\cite[Theorem~4]{vertex_coloring_induced}.


\bibliography{references}

\end{document}

%% file: table.tex
\usepackage{tikz}
\usetikzlibrary{arrows.meta,patterns}
\usetikzlibrary{ipe} 

\tikzstyle{ipe stylesheet} = [
  ipe import,
  even odd rule,
  line join=round,
  line cap=butt,
  ipe pen normal/.style={line width=0.8},
  ipe pen fat/.style={line width=1.6},
  ipe pen heavier/.style={line width=1.2},
  ipe pen thin/.style={line width=0.4},
  ipe pen ultrafat/.style={line width=2},
  ipe pen normal,
  ipe mark normal/.style={ipe mark scale=3},
  ipe mark large/.style={ipe mark scale=5},
  ipe mark small/.style={ipe mark scale=2},
  ipe mark tiny/.style={ipe mark scale=1.1},
  ipe mark Large/.style={ipe mark scale=8},
  ipe mark normal,
  /pgf/arrow keys/.cd,
  ipe arrow normal/.style={scale=7},
  ipe arrow large/.style={scale=10},
  ipe arrow small/.style={scale=5},
  ipe arrow tiny/.style={scale=3},
  ipe arrow normal,
  /tikz/.cd,
  ipe arrows, 
  <->/.tip = ipe normal,
  ipe dash normal/.style={dash pattern=},
  ipe dash dotted/.style={dash pattern=on 1bp off 3bp},
  ipe dash dash dot dotted/.style={dash pattern=on 4bp off 2bp on 1bp off 2bp on 1bp off 2bp},
  ipe dash dash dotted/.style={dash pattern=on 4bp off 2bp on 1bp off 2bp},
  ipe dash dashed/.style={dash pattern=on 4bp off 4bp},
  ipe dash densely dotted/.style={dash pattern=on 1bp off 1bp},
  ipe dash normal,
  ipe node/.append style={font=\normalsize},
  ipe stretch normal/.style={ipe node stretch=1},
  ipe stretch normal,
  ipe opacity 10/.style={opacity=0.1},
  ipe opacity 20/.style={opacity=0.2},
  ipe opacity 30/.style={opacity=0.3},
  ipe opacity 40/.style={opacity=0.4},
  ipe opacity 50/.style={opacity=0.5},
  ipe opacity 60/.style={opacity=0.6},
  ipe opacity 70/.style={opacity=0.7},
  ipe opacity 80/.style={opacity=0.8},
  ipe opacity 90/.style={opacity=0.9},
  ipe opacity opaque/.style={opacity=1},
  ipe opacity opaque,
]
\definecolor{gray2}{rgb}{0.596,0.596,0.596}
\definecolor{blue1}{rgb}{0.274,0.392,0.666}
\definecolor{blue2}{rgb}{0.529,0.607,0.784}
\definecolor{blue3}{rgb}{0.784,0.819,0.901}
\definecolor{brown1}{rgb}{0.654,0.509,0.18}
\definecolor{brown2}{rgb}{0.776,0.682,0.466}
\definecolor{brown3}{rgb}{0.898,0.854,0.752}
\definecolor{cyan1}{rgb}{0.137,0.631,0.878}
\definecolor{cyan2}{rgb}{0.439,0.76,0.921}
\definecolor{cyan3}{rgb}{0.741,0.89,0.964}
\definecolor{darkorange1}{rgb}{1,0.549,0}
\definecolor{darkorange2}{rgb}{1,0.663,0.251}
\definecolor{darkorange3}{rgb}{1,0.773,0.498}
\definecolor{darkorange4}{rgb}{1,0.867,0.702}
\definecolor{darkorange5}{rgb}{1,0.957,0.902}
\definecolor{grassgreen1}{rgb}{0.02,0.569,0.161}
\definecolor{grassgreen2}{rgb}{0.082,0.675,0.235}
\definecolor{grassgreen3}{rgb}{0.314,0.757,0.427}
\definecolor{grassgreen4}{rgb}{0.537,0.835,0.616}
\definecolor{grassgreen5}{rgb}{0.725,0.913,0.772}
\definecolor{gray1}{rgb}{0.301,0.301,0.301}
\definecolor{gray3}{rgb}{0.89,0.89,0.89}
\definecolor{green1}{rgb}{0,0.588,0.509}
\definecolor{green2}{rgb}{0.352,0.733,0.682}
\definecolor{green3}{rgb}{0.701,0.878,0.854}
\definecolor{maygreen1}{rgb}{0.549,0.713,0.235}
\definecolor{maygreen2}{rgb}{0.709,0.815,0.505}
\definecolor{maygreen3}{rgb}{0.866,0.913,0.772}
\definecolor{nicegreen1}{rgb}{0.345,0.812,0.224}
\definecolor{nicegreen2}{rgb}{0.51,0.859,0.42}
\definecolor{nicegreen3}{rgb}{0.67,0.902,0.608}
\definecolor{nicegreen4}{rgb}{0.804,0.945,0.769}
\definecolor{orange1}{rgb}{0.874,0.607,0.105}
\definecolor{orange2}{rgb}{0.921,0.745,0.419}
\definecolor{orange3}{rgb}{0.964,0.882,0.733}
\definecolor{purple1}{rgb}{0.639,0.062,0.486}
\definecolor{purple2}{rgb}{0.768,0.392,0.666}
\definecolor{purple3}{rgb}{0.894,0.717,0.847}
\definecolor{raspberry1}{rgb}{0.898,0.224,0.357}
\definecolor{raspberry2}{rgb}{0.906,0.361,0.463}
\definecolor{raspberry3}{rgb}{0.894,0.537,0.604}
\definecolor{raspberry4}{rgb}{0.894,0.761,0.788}
\definecolor{red0}{rgb}{1,0.192,0.075}
\definecolor{red1}{rgb}{0.635,0.133,0.137}
\definecolor{red2}{rgb}{0.764,0.439,0.439}
\definecolor{red3}{rgb}{0.89,0.741,0.741}
\definecolor{signalblue1}{rgb}{0.227,0.463,0.941}
\definecolor{signalblue2}{rgb}{0.42,0.562,0.941}
\definecolor{signalblue3}{rgb}{0.573,0.694,0.941}
\definecolor{signalblue4}{rgb}{0.749,0.812,0.941}
\definecolor{signalblue5}{rgb}{0.925,0.945,0.984}
\definecolor{violet1}{rgb}{0.514,0.376,0.824}
\definecolor{violet2}{rgb}{0.635,0.533,0.871}
\definecolor{violet3}{rgb}{0.753,0.686,0.91}
\definecolor{yellow1}{rgb}{0.988,0.898,0}
\definecolor{yellow2}{rgb}{0.992,0.933,0.352}
\definecolor{yellow3}{rgb}{0.996,0.968,0.701}
\definecolor{black}{rgb}{0,0,0}
\definecolor{white}{rgb}{1,1,1}
\newcommand{\resultstable}{%
\begin{tikzpicture}[ipe stylesheet, baseline, trim left]
  \fill[signalblue5]
    (254.888, 112) rectangle (395, 64);
  \fill[signalblue5]
    (150.705, 32) rectangle (254.888, 0);
  \fill[white]
    (395, 64) rectangle (254.888, 0);
  \fill[white]
    (395, 208) rectangle (254.888, 112);
  \fill[signalblue4]
    (71.5461, 208.081)
     -- (71.5461, 0.081)
     -- (150.705, 0.081)
     -- (150.705, 32.081)
     -- (254.888, 32)
     -- (254.888, 208)
     -- cycle;
  \node[ipe node]
     at (92, 217.35) {$ \subseteq P \boxtimes P$};
  \node[ipe node]
     at (177.894, 217.339) {$ \subseteq_{\text{ind}} P \boxtimes P$};
  \node[ipe node]
     at (309.824, 217.339) {$G \boxtimes P_\ell$};
  \node[ipe node, font=\small]
     at (85.462, 133.838) {\NP-complete};
  \node[ipe node, font=\small]
     at (177.133, 133.757) {\NP-complete};
  \node[ipe node, font=\small]
     at (274.203, 195.836) {\FPT{} in clique-width of $G$};
  \node[ipe node, font=\small]
     at (293.538, 163.258) {\P{} for $G$ chordal};
  \node[ipe node, font=\small]
     at (279.329, 131.836) {\P{} for $G$ co-comparable};
  \node[ipe node, font=\small]
     at (85.462, 51.917) {\NP-complete};
  \node[ipe node, font=\small]
     at (177.133, 51.836) {\NP-complete};
  \node[ipe node, font=\small]
     at (85.462, 19.917) {\NP-complete};
  \node[ipe node, anchor=north west, font=\small]
     at (156.322, 26.066) {\textsc{$3$-EC} on planar graphs};
  \node[ipe node, anchor=north west, font=\small]
     at (312.775, 19.113) {trivial};
  \node[ipe node, font=\small]
     at (290.848, 51.836) {\P{} for $G$ bipartite};
  \node[ipe node]
     at (4.846, 44.596) {\textsc{Max Cut}};
  \node[ipe node, anchor=north, font=\small]
     at (111.126, 44.242) {\cref{thm:max-cut-hardness-subgraphs}};
  \node[ipe node, anchor=north, font=\small]
     at (202.797, 44.161) {\cref{thm:max-cut-hardness-induced}};
  \node[ipe node, anchor=north, font=\small]
     at (324.944, 44.161) {\cref{cor:maxcut_bipartite}};
  \node[ipe node, anchor=north, font=\small]
     at (324.944, 188.161) {\cref{thm:fvs_cliquewidth}};
  \node[ipe node, anchor=north, font=\small]
     at (324.944, 156.742) {\cref{thm:fvs_chordal}};
  \node[ipe node, anchor=north, font=\small]
     at (324.944, 124.161) {\cref{thm:fvs_cocomparability}};
  \node[ipe node, anchor=north, font=\small]
     at (202.797, 12.161) {\cref{thm:edge-coloring-main-thm}};
  \node[ipe node, anchor=north, font=\small]
     at (111.126, 12.242) {\cref{thm:edge-coloring-main-thm}};
  \draw
    (0, 208)
     -- (395, 208);
  \draw
    (0, 32)
     -- (395, 32);
  \draw[gray2, ipe pen thin]
    (254.888, 176)
     -- (395, 176);
  \draw[gray2, ipe pen thin]
    (254.888, 144)
     -- (395, 144);
  \draw[gray2, ipe pen thin]
    (324.944, 112)
     -- (324.944, 64);
  \node[ipe node, anchor=base, font=\small]
     at (111.126, 120.081) {\cite{fvs_induced_grid}};
  \node[ipe node, anchor=base, font=\small]
     at (202.797, 120) {\cite{fvs_induced_grid}};
  \begin{scope}[shift={(-0.1994, -401.702)}]
    \node[ipe node]
       at (5.045, 420.596) {\textsc{$3$-Edge}};
    \node[ipe node]
       at (5.045, 408) {\textsc{Coloring}};
  \end{scope}
  \begin{scope}[shift={(4.8455, -401.702)}]
    \node[ipe node]
       at (0, 540.596) {\textsc{Feedback}};
    \fill[gray3]
      (16, 528)
       -- (16, 528);
    \node[ipe node]
       at (0, 528) {\textsc{Vertex Set}};
  \end{scope}
  \begin{scope}[shift={(-21.9998, -408)}]
    \node[ipe node, font=\small]
       at (298.235, 506.418) {$\ell$ even:};
    \node[ipe node, font=\small]
       at (293.33, 492.937) {\textsc{VC} on $G$};
    \node[ipe node, anchor=north, font=\small]
       at (311.916, 485.582) {Cor.\ \ref{cor:vertex_cover}};
  \end{scope}
  \begin{scope}[shift={(-9.9999, -408)}]
    \node[ipe node, font=\small]
       at (357.702, 506.443) {$\ell$ odd:};
    \node[ipe node, font=\small]
       at (345.249, 492.887) {\textsc{$\ell$-FVS} on $G$};
    \node[ipe node, anchor=north, font=\small]
       at (369.972, 485.557) {Obs.\ \ref{obs:ell_weighted_forest}};
  \end{scope}
  \draw
    (0, 64)
     -- (395, 64);
\end{tikzpicture}%
}